%% file: main.tex
\documentclass[acmlarge]{acmart}
\makeatletter
\newcommand{\myconfshort}{\acmConference@shortname}
\newcommand{\myconffull}{\acmConference@name}
\newcommand{\myconfdate}{\acmConference@date}
\newcommand{\myconfloc}{\acmConference@venue}
\AtBeginDocument{
  \fancypagestyle{firstpagestyle}{
    \fancyhead{}%
    \fancyfoot[C]{}%
  }
  \fancyhf{}
  \fancyhead[LO]{\@headfootfont\shorttitle}%
  \fancyhead[RE]{\@headfootfont\@shortauthors}%
  \fancyhead[LE]{\@headfootfont\footnotesize \myconfshort, \myconfdate, \myconfloc}%
  \fancyhead[RO]{\@headfootfont\footnotesize \myconfshort, \myconfdate, \myconfloc}%
  \fancyfoot[C]{}%
}
\makeatother
\acmBooktitle{\conffull\@ (\confshort), \confdate, \confloc}
\AtBeginDocument{%
  }

\copyrightyear{2026}
\acmYear{2026}
\setcopyright{cc}
\setcctype{by}
\acmConference[FAccT '26]{The 2026 ACM Conference on Fairness, Accountability, and Transparency}{June 25--28, 2026}{Montreal, QC, Canada}
\acmBooktitle{The 2026 ACM Conference on Fairness, Accountability, and Transparency (FAccT '26), June 25--28, 2026, Montreal, QC, Canada}
\acmDOI{10.1145/3805689.3812369}
\acmISBN{979-8-4007-2596-8/2026/06}

\usepackage[dvipsnames, table]{xcolor}
\usepackage{booktabs}
\usepackage{tabularx}
\usepackage{adjustbox}
\usepackage{array}
\usepackage{multirow, multicol}
\usepackage{subcaption}
\usepackage{amsmath}
\usepackage{tabto}
\usepackage{verbatim}
\usepackage{dcolumn}
\usepackage{flafter} 
\usepackage{soul}

\begin{document}
\input{template/commands}

\title{
Framing an AI with Values Reduces AI Reliance in AI-supported Writing Tasks
}

\author{Alice Gao}
\affiliation{%
  \institution{University of Washington}
  \city{Seattle}
  \country{USA}}
\email{atgao@cs.washington.edu}

\author{Andrew N. Meltzoff}
\affiliation{%
  \institution{University of Washington}
  \city{Seattle}
  \country{USA}}
\email{meltzoff@uw.edu}

\author{Maarten Sap}
\affiliation{%
  \institution{Carnegie Mellon University}
  \city{Pittsburg}
  \country{USA}}
\email{msap2@andrew.cmu.edu}

\author{Katharina Reinecke}
\affiliation{%
  \institution{University of Washington}
  \city{Seattle}
  \country{USA}}
\email{reinecke@cs.washington.edu}

\renewcommand{\shortauthors}{Gao et. al}

\input{files/00_abstract}

%
%



\maketitle
\input{files/01_intro}
\input{files/02_rw}
\input{files/03_methods}
\input{files/04_findings}

\input{files/05_discussion}
\input{files/06_limitations}
\input{files/07_conclusion}

\bibliographystyle{ACM-Reference-Format}
\bibliography{references}
\clearpage

\appendix
\input{files/90_study}
\input{files/91_demographics}
\input{files/92_results}
\input{files/93_addresults}
\input{files/94_suggestionvals}

\end{document}

%% file: template/commands.tex
\newcommand{\todo}[1]{\textcolor{Red}{\textbf{TODO: \textit{#1}}}}
\newcommand{\todocite}[1]{\textcolor{Orange}{\textbf{\textit{CITE: #1}}}}
\newcommand{\placeholder}{\textcolor{cyan}{\textbf{\textit{[...]}}}}

\newcommand{\rev}[1]{\textcolor{blue}{#1}}
\newcommand{\del}[1]{\textcolor{red}{\st{#1}}}

\newcommand{\ag}[1]{\textcolor{Purple}{\textbf{AG: #1}}}
\newcommand{\kr}[1]{\textcolor{olive}{\textbf{KR: #1}}}
\newcommand{\ms}[1]{\textcolor{teal}{\textbf{MS: #1}}}
\newcommand{\am}[1]{\textcolor{green}{\textbf{AM: #1}}}

%% file: files/00_abstract.tex
\begin{abstract}
Despite a global user base adopting large language models (LLMs) for daily writing tasks, model suggestions tend to align with Western values. Research has shown users commonly accept a high fraction of these AI suggestions, homogenizing writing styles and rendering outputs more ``Western'' than intended. While this suggests a need to reduce AI reliance, it remains unknown what kind of interventions could achieve this. Can framing the AI with specific values, and comparing it to one's own, make users less susceptible to overreliance and support more unique writing? We tested this hypothesis in a between-subjects online experiment with Indian and American participants (n=149) in which they were asked to perform AI-supported writing tasks, either 1) without an intervention, 2) after seeing an overview of the AI's framed values, or 3) after seeing an overview of the AI's framed values compared to their own. Our results show that seeing the AI's framed values reduces AI reliance, i.e., the proportion of the final essay generated by the AI, by an average of 20\%. Additionally, when participants saw an overview of the AI's framed values (without comparison to their own values), the final essays contain more unique text than without intervention. Our findings emphasize the importance of educating users about potential value biases in AI, showing that raising awareness with a simple overview of values encourages users to personalize their writing.  
\end{abstract}

%% file: files/01_intro.tex
\section{Introduction} \label{sec:intro}
Large language models (LLMs) have become one of the most widespread AI applications with many people using it to fully generate text or autocomplete sentences in emails \cite{kannan2016smartreply}, scientific writing \cite{gero2021sparksinspirationsciencewriting}, or journalism \cite{petridis2023anglekindling}. Despite a global uptake of AI-supported writing systems, models remain primarily trained on English language and Western-centric data~\cite{johnson2022ghostmachineamericanaccent, bender2021stochastic, bella2024tackling}, frequently leading their outputs to align more with Western values~\cite{santy2023nlpositionalitycharacterizingdesignbiases}. Research has found that users of AI-supported writing systems can perceive these suggestions as inadequately representing their culture, language, and sociolect, feeling that these systems were not designed with them in mind~\cite{basoah2025cscw}. AI suggestions have also been shown to change a users' language \cite{hohenstein2023} to produce more generically positive outputs \cite{arnold2018predictive, arnold2018bias, hohenstein2023}, shifting their attitudes about social topics~\cite{jakesch2023cowriting}, and influencing which topics individuals write about \cite{poddar2023influence}. Moreover, because people tend to accept a high fraction of AI suggestions, the output of AI-supported writing tends to be in line with the AI's value biases, leading to more homogenized, generic outputs than if people were not using AI~\cite{agarwal2025, arnold2018predictive, anderson2024homogenization}. 

This value mismatch can cause harms during human-AI interaction, as the AI may continue to promote values at odds with users' diverse experiences, leading to the loss of agency or even contribute to the erasure of unique cultures, languages, or sociolects~\cite{ghosh2023chatgpt, basoah2025cscw}. This homogenization remains an issue as it may erode distinct cultural practices and creative expressions, such as when AI is used in story-writing interfaces \cite{lee2022coauthor, singh2023story, yang2022AIAA}, or alter cultural preferences and reinforce Western values. The human tendency to engage in \textit{satisficing}, i.e., choosing the first reasonable option presented because it saves time and cognitive effort to do so \cite{simon1971human, northcraft1990organizational}, further exacerbates the potentials of these harms during AI-supported writing. 

Satisficing has also been demonstrated in AI-supported decision making \cite{bansal2019beyond, bucina2021trust, todd1994decisionaid}, where users tend to overrely on AI systems to distinguish correct and wrong answers. While intervention strategies have been developed and shown to successfully reduce overreliance in AI-supported decision making, such as providing explanations \cite{vasconcelos2023overreliance} or nudging users to engage analytically with an AI before they can make their final decision~\cite{bucina2021trust}, we have limited knowledge of what types of interventions can reduce overreliance or overacceptance of AI suggestions during AI-supported writing. 
In this work, we address this gap by evaluating whether simply making users aware of an AI's potential values can make them less susceptible to relying on the AI's suggestions and lead to more diverse and unique texts. Our main research question asks: \textbf{Does framing an AI writing assistant with specific values affect user reliance and the extent to which they add their own unique words to the output?}

In a between-subjects online study with 149 participants from India and the US, we evaluated this question by asking participants to perform two co-writing tasks with an AI\footnote{Though our study specifically investigates LLMs and their influence on writing, we framed our intervention to participants as potential values an AI may have. Furthermore, lay users typically think of AI more broadly instead of LLMs. As such we use the terms interchangeably throughout the paper.} with two types of interventions, a non-comparative and comparative overview of the framed AI values. 
We tested these interventions across four conditions: (1) no intervention (baseline), (2) after seeing an overview of the framed AI values (AI values), (3) after considering their own values and seeing an overview of the framed AI values (primed + AI Values), or (4) after considering their own values and seeing an overview the framed AI values in comparison to their own (primed + comparative AI values). 

Our results indicate that providing an overview of specific AI values reduces user reliance on the AI by an average of 20\% and leads to greater uniqueness in responses, but only when participants did not see these values in comparison to their own. In line with prior work~\cite{agarwal2025}, our Indian participants generally had a much higher AI reliance and accepted more AI suggestions than Americans. However, as we hypothesized, seeing the framed AI's values resulted in lower AI reliance compared to the baseline for both Indian and American participants. As a result, the output text tended to be more diverse and unique after seeing the intervention, suggesting that awareness of potential value biases (in a non-comparative overview) encourages users to put more effort into contributing their own unique words. We conclude by discussing the implications of designing effective intervention strategies during AI co-writing and the importance of surfacing an AI's value biases to users. 

%% file: files/02_rw.tex
\section{Related Work}
We contextualize our work by first providing an overview of AI overreliance and mitigation strategies. We then provide a background of the biases and values within AI and the influence of these biases.

\subsection{Reliance and Interventions}
As LLMs and AI assistants have become integrated into multiple aspects of our lives, users have also become increasingly \textit{reliant} on LLMs, accepting automated decisions as a means of reducing cognitive effort and saving time, or satisficing \cite{todd1994decisionaid}, even when the LLM is wrong \cite{fogliato2022hai, bucina2021trust}. Researchers have studied and designed potential strategies to reduce overreliance that arises due to satisficing, particularly during AI assisted decision-making, such as providing explanations at how an LLM arrived at its decisions \cite{bayati2014readmissions, vasconcelos2023overreliance}, indications of the AI's certainties \cite{bansal2021doesexceedpartseffect, lai2020deception}, delaying the delivery of AI recommendations \cite{park2019slow}, or asking individuals to first make a decision before engaging with the AI's decision \cite{fogliato2022hai, green2019algoinloop}.

These studies reveal a complex relationship between interventions and reducing overreliance: explanations of an AI's decisions or recommendations alone do not reduce overreliance \cite{bucina2021trust, vasconcelos2023overreliance, bansal2021doesexceedpartseffect}. In fact, explanations can actually increase overreliance as their presence may increase trust in LLMs \cite{zhang2020effect, yun2019trust}. Instead, individuals must be forced to actively engage with these explanations \cite{bucina2021trust}, motivated by external benefits \cite{vasconcelos2023overreliance}, or view engaging with LLM explanations as less difficult than the task at hand \cite{vasconcelos2023overreliance}. 
To add to this complexity, prior work has repeatedly found that the context and framing of AI explanations can increase reliance. Using more anthropomorphic language \cite{zhou2025relai} or expressing confidence in responses \cite{rathi2025humansoverrelyoverconfidentlanguage} increases user reliance and trust, as does the framing of the AI assistant as an expert or companion part of one's group \cite{basoah2025anthromorphic, chen2026presentinglargelanguagemodels} during human-AI interactions. This suggests that reducing overreliance is a complicated task that must be approached with care, taking the user's prior experience and perceived effort as well as the framing of the interaction with the AI into account. 
Most overreliance-reducing strategies focus on reducing AI reliance during \textit{decision-making}, where users distinguish correct and incorrect choices, and less so when individuals collaborate with embedded or interactive AI tools where they seek LLM outputs to be more productive, producing writing for work or for personal tasks, requiring uninterrupted flow. Our work seeks to fill this gap by building upon existing work in reducing overreliance through developing a non-disruptive intervention with interactive AI tools during AI-supported writing.

\subsection{Value Biases and the Influence of AI}
Despite the pervasiveness of LLM-powered tools, these models embed the biases of their training data, which is frequently Western-centric \cite{bella2024tackling, johnson2022ghostmachineamericanaccent}. As a result, these tools are limited in their ability to serve diverse user groups \cite{bella2024tackling}, and produce better results for well-represented groups \cite{wagner2021wikipedia, lee2024portray}. Even with guardrails to remove explicit biases, the outputs of LLMs are embedded with \textit{implicit} biases learned from their data \cite{bianchi2023t2i, yuxuan2025actions}. Prior work has shown how LLMs flatten nuances into homogeneous outputs \cite{lee2024portray}, portraying racial minorities as one-dimensional or providing a Western-centric lens for these groups \cite{agarwal2025, lee2024portray}. Other work has shown how these biases can lead to minorities feeling excluded and unaccounted for when LLMs fail to recognize diverse cultures, languages, or sociolects \cite{basoah2025cscw, gadiraju2023offensive, bella2024tackling, bender2021stochastic, gehman2020realtoxicityprompts} resulting in poor user experience and frustration \cite{chen2024conversational, lee2024portray, bella2024tackling}. 

Researchers have studied the implications of using these tools as their ubiquity has increased. Work has shown how AI suggestions can influence a user's language \cite{hohenstein2023} and result in generic, and overly positive text \cite{arnold2018predictive, arnold2018bias}, while \citet{agarwal2025} showed that Indian participants who used AI-supported writing produced more homogenized writing, lacking nuances and portraying cultural artifacts from a \textit{Western} lens. \citet{jakesch2023cowriting} also showed how writing suggestions from opinionated models are a form of \textit{latent persuasion} that can shift users' social attitudes. 

Though work has shown simply informing people of biases can mitigate their influence \cite{pope2018awareness, lee2017awareness, guynn2015google}, this problem becomes much more complex with LLMs with their known inconsistencies \cite{rottger2024politicalcompassspinningarrow}. LLMs, just like humans, are inconsistent with their values: the values a person may indicate may not be the values they adhere to in their own actions and tasks \cite{vermeir2006, sweeny2012say} and similarly, probabilistic models are inconsistent in value expression in value-laden situations \cite{moore2024largelanguagemodelsconsistent, khan2025random}. Taking this into account, our work investigates the impact of framing an AI writing assistant with specific values rather than attempting to elicit values that may be inconsistent. 

%% file: files/03_methods.tex
\section{Methods}
We conducted a controlled online experiment to determine whether framing an AI with specific values affects participants' interaction with the LLM and the generated output. We adapted ~\citet{agarwal2025}'s study which found that Indian and American participants heavily rely on LLM writing suggestions. We added the following interventions to test whether showing participants an AI's framed values could reduce this reliance: 
\begin{enumerate}
    \item \textbf{No Intervention (baseline):} Our baseline where participants do not see an AI's framed values. 
    \item \textbf{AI Values:} Participants view an AI's framed values before the writing tasks and values survey. 
    \item \textbf{Primed + AI Values:} Participants take the values survey \textit{then} view an AI's framed values before completing the writing tasks.
    \item \textbf{Primed + Comparative AI Values (comparative):} Participants take the value survey then view their values \textit{compared} to an AI's framed values before completing the writing tasks. 
\end{enumerate}
We manipulated the type of overview (non-comparative vs. comparative) participants saw as we hypothesized that a comparison provides better understanding \cite{festinger1954theory} of the AI values shown. People often draw comparisons with relevant individual factors \cite{fleischmann2021more, wu2012brain, boecker2022individuals} to process information in the world. We hypothesized that an increased understanding would decrease participants' reliance. We tested two timings of the non-comparative overview, primed + AI values and AI values, as the former may cause participants to implicitly compare their own values to the framed AI values. Table \ref{tab:study-flow} provides an overview of our experimental procedure.

\begin{table*}[!htpb]
    \small
    \centering
    \caption{Study flow for four participant groups. No intervention, which is our baseline, where they did not view the AI's framed values. AI values and primed + AI both saw the AI's framed values but in the former they viewed it before the writing tasks. Primed + Comparative AI values (comparative) condition participants viewed their values in comparison to the framed AI values shown.}
    \begin{tabular}{lp{.2\textwidth}p{.25\textwidth}p{.3\textwidth}}
    \toprule
         \textbf{No Intervention} & \parbox{.25\textwidth}{\textbf{AI Values}} & \parbox{.25\textwidth}{\textbf{Primed + AI Values}} & \parbox{.3\textwidth}{\textbf{Primed + Comparative AI Values}}\\  
         \noalign{\smallskip}
         \midrule
        1. Demographic Survey & 1. Demographic Survey & 1. Demographic Survey & 1. Demographic Survey \\ 
        2. \textit{\textbf{Values Survey}}& 2. \textbf{AI Value Overview} & 2. \textit{\textbf{Values Survey}} & 2. \textit{\textbf{Values Survey}} \\ 
        3. Writing Task 1 & 3. Writing Task 1 & 3. \textbf{AI Value Overview} & 3. \textbf{Comparative AI Value Overview} \\ 
        4. Writing Task 2 & 4. Writing Task 2 & 4. Writing Task 1 & 4. Writing Task 1 \\ 
        5. AI Literacy Survey & 5. \textit{\textbf{Values Survey}} & 5. Writing Task 2 & 5. Writing Task 2 \\
        & 6. AI Literacy Survey & 6. AI Literacy Survey & 6. AI Literacy Survey \\ 
        \bottomrule
    \end{tabular}
    \label{tab:study-flow}
\end{table*}

\subsection{Hypotheses}
We established a set of hypotheses to answer our main research question: \\
\noindent 
\noindent\textbf{Hypotheses 1: Effect on AI Reliance.} 
We hypothesized that seeing an AI's framed values would increase user understanding that an AI can have values and biases, and that this understanding would reduce AI reliance and acceptance of its suggestions. Furthermore, we hypothesized that greater understanding of the difference between an AI's framed values and one's own would lead to even further decreases of AI reliance.
\begin{quote}
    \textbf{H1a:} Users that see an AI's framed values (AI values, primed + AI values, comparative) will have a lower AI reliance and AI acceptance rate compared to users that do not see an AI's framed values (baseline). \\
    \textbf{H1b:} Users that see an AI's framed values in \textit{comparison} to their own (comparative) will have a lower AI reliance and AI acceptance rate compared to users that just see an AI's framed values (AI values, primed + AI values). 
\end{quote}
Additionally, we hypothesized that the greater the value difference between a participant and the selected AI values, the lower their AI reliance and AI acceptance rate: 
\begin{quote}
    \textbf{H1c:} There exists a negative correlation between an individual's value difference with an AI's framed values and their AI reliance and AI acceptance rate. 
\end{quote}

\noindent\textbf{Hypotheses 2: Effect on Uniqueness.} 
We also hypothesized that framing an AI with specific values and an increased understanding would lead users to making their texts more unique, or contributing more of their own words. The understanding would implicitly promote users to counteract these biases through individualizing their responses. We also hypothesized that greater understanding would further increase this uniqueness. 
\begin{quote}
    \textbf{H2a:} Users that see the AI's framed values (AI values, primed + AI values, comparative) will produce more unique texts compared to users that do not see the AI's framed values (baseline). \\
    \textbf{H2b:} Users that see the AI's framed values in \textit{comparison} to their own (comparative) will produce more unique texts compared to users that just see the AI's framed values (AI values, primed + AI values). 
\end{quote}

Because prior work has shown that Indian users tend to have a higher reliance on AI than American users~\cite{agarwal2025}, we additionally explored whether the effect of the interventions differed between the two participant groups. 

\subsection{Recruitment \& Participants}
Participants were recruited from Prolific, an online platform tailored for academic research \cite{palan2018prolific}, enabling us to target specific participant groups. We recruited participants over 18, who had English fluency, and lived either in the United States, for Americans, or India, for Indians, and nowhere abroad for more than 6 months. A priori power analysis conducted using \texttt{R}'s \texttt{pwr} package and indicated a minimum of 31 participants per group for four groups within a moderate effect size (0.3), at a power of 0.80 and $\alpha$ of 0.05. We aimed to recruit 40 participants per condition with an equal number of Indian (n=20) and American (n=20) participants but only analyzed the data of participants that provided valid responses. Responses were considered invalid if they did not answer the correct writing prompts. In total, we had 149 participants with valid responses: 40 participants formed the baseline group, 39 participants in AI values condition, 34 participants in primed + AI values condition, and 36 participants in comparative condition. 76 participants were from the United States and 73 participants were from India. We provide the full list of self-reported participant demographics in Table \ref{tab:demographics}. 

The study protocol was reviewed and received exempt status by the first author's university's Institutional Review Board (IRB). Participants were compensated at an hourly rate of \$12 an hour for study completion. 

\subsection{Study Procedure}
Participants were redirected to our custom study portal, built using SvelteKit and hosted at a public URL. Upon entering our study portal, participants first confirmed their consent and provided their unique Prolific participant ID. They then took a short demographic survey. After, participants proceeded to either a 15 question values survey or saw the AI's framed values. Table \ref{tab:study-flow} provides the experimental flow for all four conditions.

Following Agarwal et al.'s procedure \cite{agarwal2025}, participants were required to complete two writing tasks, shown sequentially, with a 50 word minimum. Once participants reached the minimum word requirement, they could proceed to the next step of the study. They were shown their word count in the bottom as they typed. To encourage participants to not solely rely on the LLM's suggestions, the initial Prolific recruitment framed the goal of the study as a way to collect personal stories from around the world. We show our interface in Fig. \ref{fig:writing-task-portal}.

All participants finished with a four question AI literacy survey. Upon completion, they were provided a code they could enter into Prolific to verify their completion and receive monetary compensation. All survey answers, final written responses, and writing task interaction logs were all logged to an external database.

\subsection{Materials}
\subsubsection{Intervention Strategy: an AI Value Overview}\label{sec:intervention}
We created an \emph{AI value overview} to test whether users change their propensity to accept AI suggestions once they have seen an \textit{AI's framed values.} To do so, we first required values for the AI. Although we considered using the Short Schwartz Values Survey \cite{lindeman2005ssvs} or Inglehart-Welzel's 10 questions \cite{welzel2010agency} from the World Values Survey (WVS) \cite{wvs_wave7}, we chose to manually selected a subset of questions from the WVS to circumvent the Western biases and criticism Eurocentric value constructs found in these questions \cite{banerjee2018interpretation, bomhoff2012easiadiff}, as well as their limited applicability for other cultural groups \cite{bomhoff2012easiadiff}. As the WVS is over 200 questions, we selected a subset of 15 questions across its 14 thematic subsections based on concreteness, understandability, and ability to provide meaningful comparison to a user's own values. These questions were chosen by the first author then discussed and iterated upon in meetings with the entire team until consensus. Our questions span four value dimensions: social, trust, security, and religion. These value dimensions were obtained through the WVS thematic category that questions were selected from. To obtain the ``AI's values,'' we used OpenAI's \texttt{GPT-4o} as our ``AI'' and queried it on our subset of questions and answer choices for at least 10 trials per question. The average of Likert questions and the majority answer for multiple choice questions were taken as the AI's framed values. In case of ties, we queried the AI an additional time. The full list of questions can be found in Appendix \ref{wvs}. 

\begin{figure*}[!htbp]
    \centering
    \includegraphics[width=0.9\textwidth]{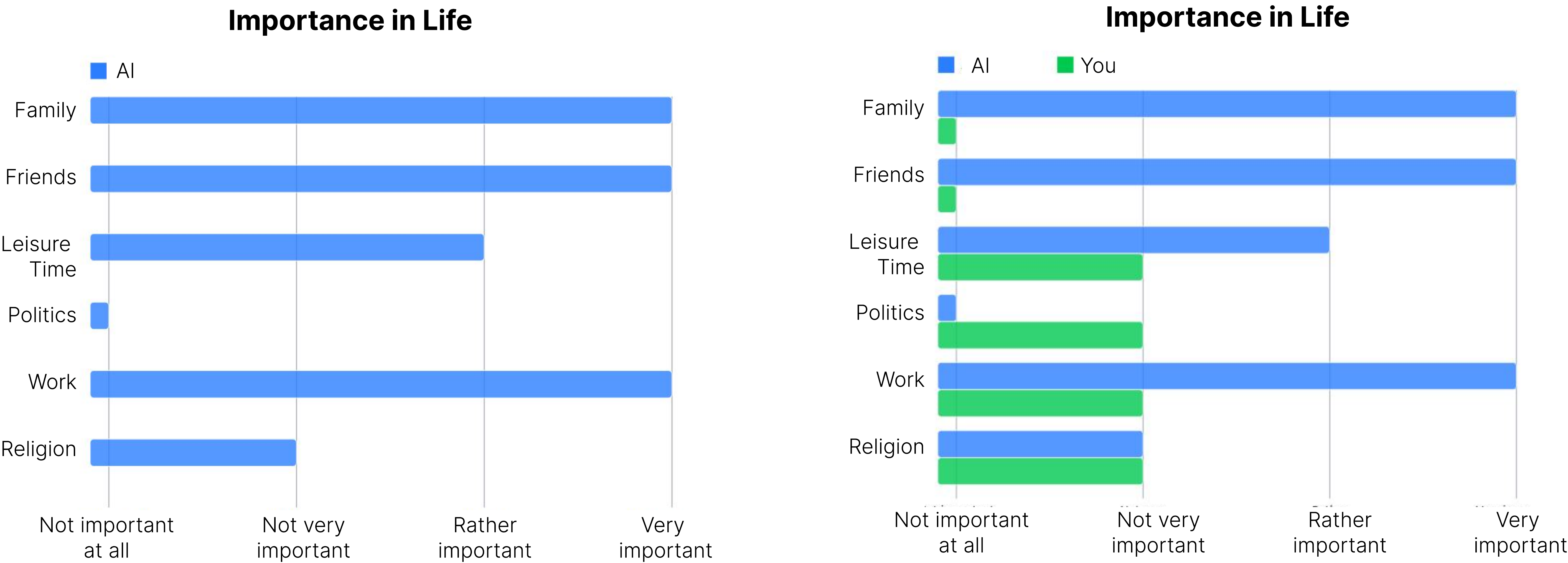}
    \caption{A snapshot of our intervention, showing the AI's framed values we showed participants with the AI's answer to ranking six items (family, friends, leisure time, politics, work, and religion) on the importance in its life. The left is shown to the non-comparative interventions (AI values and primed + AI values) while the right shows the participant's values compared to the AI's (comparative).}\label{value-profile}
\end{figure*}

We provided the specified values as a bar chart as it is a common, easy to read visualization \cite{kosslyn1989charts}. The same AI value overview was shown to all participants. Fig. \ref{value-profile} shows a portion of this overview and the comparative version, where participants saw their values compared to the AI's. We added two conditions that showed the intervention at different points in the study as seeing the AI's framed values after taking the values survey could cause indirect comparison of participants' own values to the framed values. 

Though prior work has characterized the biases embedded in AI and identified potential misalignments with different user groups \cite{santy2023nlpositionalitycharacterizingdesignbiases, tao2024culturalbias}, capturing an AI's ``values'' remains an open question \cite{shiffrin2023probing, khan2025randomness}. Our AI value overview was not designed to perform the impossible task of fully reflecting the AI's values, but to signal to users that an AI had these specified values and frame their interaction with the AI through the lens of a value-laden AI. It is possible that the AI suggestions did not align with the values in our overview. As a way to assess this, we performed an additional analysis to determine the overlap between the suggestions shown in the study and the framed AI values (Appendix \ref{app:suggestion-vals}). We found that a majority of suggestions were neutral towards the framed AI values. We further discuss the limitations of our approach in \S\ref{sec:discussion} and \S\ref{sec:limits}.

\subsubsection{Writing Task}
Each participant completed two writing tasks randomly chosen from two categories: (1) writing tasks where AI suggestions tend to exhibit Western bias and (2) writing tasks where AI suggestions exhibit less Western bias. These tasks were adapted from Agarwal et. al's \cite{agarwal2025}, meant to elicit explicit cultural practices and more nuanced values. For the first category, participants wrote about their favorite food or holiday. For the second category, participants wrote about their favorite celebrity/public figure or asked their boss for a two-week leave. We provide all prompts in Table \ref{tab:writing-tasks}. For each writing task, participants were able to accept or reject AI autocomplete suggestions.

We adopted Agarwal et. al's \cite{agarwal2025} procedure to retrieve autocomplete suggestions from \texttt{GPT-4o}, the latest OpenAI model at the time. Autocomplete suggestions were shown as light gray ghost text, similar other autocomplete suggestions text-based editors like Gmail, and limited to 10 words. We embedded the essay topic in the prompt and the participant's current input into the prompt to provide contextually relevant suggestions \cite{buschek2021impact, jakesch2023cowriting, agarwal2025}. Suggestions were retrieved every 500ms, like in prior work \cite{buschek2021impact, jakesch2023cowriting}, and could be accepted by pressing the \texttt{Tab} button or ignored by continuing to type. To prevent participants from rapidly accepting suggestions we required that they type another letter after accepting an AI's suggestion. The full prompt to retrieve suggestions is provided in Table \ref{writing-suggestion-prompt}.

\subsection{Analysis}\label{sec:metrics}
We analyzed three sets of data: interaction logs captured on the study portal, the final essays written by the participants, and personal factors. We summarize the metrics used in our analyses below. 

\subsubsection{AI Reliance}
From writing task interaction logs, we used following metrics to assess an individual's AI reliance:
\begin{enumerate}
    \item \textbf{AI Reliance:} The proportion of the final essay generated by the AI \cite{buschek2021impact, kadoma2024aicomm, agarwal2025}, computed as the ratio of total number of AI-suggested characters in the final essay to the total number of characters in the essay. The longest common subsequence (LCS) was used to handle scenarios in which a suggestion was modified. It ranges from 0 to 1, where a higher value indicates higher AI reliance or more AI-generated content. This metric is sensitive to suggestions that were accepted but later modified or deleted.
    \item \textbf{AI Acceptance Rate:} The proportion of accepted AI suggestions to the total number of AI suggestions shown in a task \cite{buschek2021impact, agarwal2025} and a metric of engagement with the AI. It ranges from 0 to 1 and is a coarse measurement as users can later delete or modify suggestions. 
    \item \textbf{AI Suggestion Modification Rate:} The proportion of AI accepted suggestions that were modified to the total number of AI suggestions \cite{buschek2021impact}. It ranges from 0 to 1, and a higher number suggests that the users tended to modify AI suggests to better suit individual needs. To capture modification, we checked if an accepted suggestion appeared in the final essay. 
\end{enumerate}
We included AI suggestion modification rate in our analysis following prior work \cite{dhillon2024cowriting, buschek2021impact, agarwal2025} though it is not formally in our hypotheses as this metric is partially dependent on the AI acceptance rate.

\subsubsection{Writing Uniqueness}
We used methods from NLP literature and qualitative analysis to analyze uniqueness: 
\begin{enumerate}
    \item \textbf{Lexical Diversity:} We used type-token ratio (TTR) as a coarse measurement of lexical diversity in text \cite{baker2006glossary}, calculated as the proportion of the number of unique words (types) to the total number of words (tokens). The metric is bounded from 0 to 1 where 0 represents more repetition and less linguistic diversity. TTR has the weakness that it penalizes longer texts.
    \item \textbf{Cosine Similarity:} We retrieved embeddings from OpenAI's \texttt{text-embedding-large} model to calculate cosine similarity scores between pairs of essays on the same topic with all other users. An individual's similarity score was computed by computing cosine similarities to all same-topic essays with other participants within their condition. Each participant wrote two essays so the scores were averaged for a final similarity score. Similarities ranged from -1 to 1, where -1 represents completely opposite essays and 1 represents essays that are exactly the same. This provides a rough measurement for textual similarity but fails to consider the context of words or sentences.
    \item \textbf{Thematic Analysis:} We conducted a thematic analysis \cite{braunclark06themes} on participant essays as computational metrics provide coarse-grained measures to analyze written text. AI-supported writing is known to homogenize text \cite{hohenstein2023}, making it more positive \cite{arnold2018bias}, so we seek to capture nuances undetected through automatic metrics. 
\end{enumerate} We do not report on writing productivity, though it was collected in previous studies \cite{dhillon2024cowriting, agarwal2025}, as it not indicative of writing uniqueness.

\subsubsection{Personal Factors}
We collected two measures for an individual's personal factors: 
\begin{enumerate}\label{metric:ailit}
     \item \textbf{AI Value Difference:} We calculated the value difference between a participant's own values and the AI's framed values from our overview: 
    \begin{align*}
        \text{Value Difference} &= \frac{\sum_{A=0}^{n=15} |A_{user} - A_{AI}|}{\max \sum_{A=0}^{n=15} |A_{user} - A_{AI}|}
    \end{align*}
    where $A_{user}$ and $A_{AI}$ are the user and AI's answer to the current question respectively. Non-Likert questions were encoded binary variables where 1 represented the AI's answer and occurrence of this answer and 0 otherwise. The value difference was normalized to the range of 0 (the same) to 1 (most dissimilar) for better interpretability. We used this as an exploratory measure due to the difficulties in capturing an AI's framed values, as their approach to answering questions is drawn from their training data \cite{shiffrin2023probing, frank2023baby}. Furthermore, AI responses have not been shown to correlate to real-world behavior unlike human responses to these surveys \cite{aycinea2022norms}. We collected the measure for all participants but only examined it for those that received an intervention.

    In our exploratory analysis, we further break down the overall value difference into 4 dimensions (\textit{social, trust, security, and religion)} based on the thematic section from the WVS they were selected from. 
    \item \textbf{AI Literacy:} We also collected an individual's AI literacy level and reported on this metric, as this has been shown to has been shown to impact their perception of AI \cite{mun2025whyNotUseAI}. Our AI literacy survey was adopted from prior work \cite{mun2025whyNotUseAI, mun2024participAI, wang2023measure}. This survey measures a participant's AI literacy on a 7-point Likert scale and captures a participant's AI literacy across 3 dimensions: \textit{awareness} (participants' awareness of AI in technology and ability to identify them), \textit{usage} (how frequently participants use AI for work and non-work tasks), and \textit{ethics} (participants' awareness of the limitations and shortcomings of AI technology).
\end{enumerate}

%% file: files/04_findings.tex
\section{Findings}\label{sec:findings}
\subsection{Framing an AI with specific values reduces an individual's AI reliance}\label{sec:ai-engagement}
\begin{figure*}[!hptb]
    \centering
    \begin{subfigure}[t]{0.3\textwidth}
        \centering
        \includegraphics[height=0.9in]{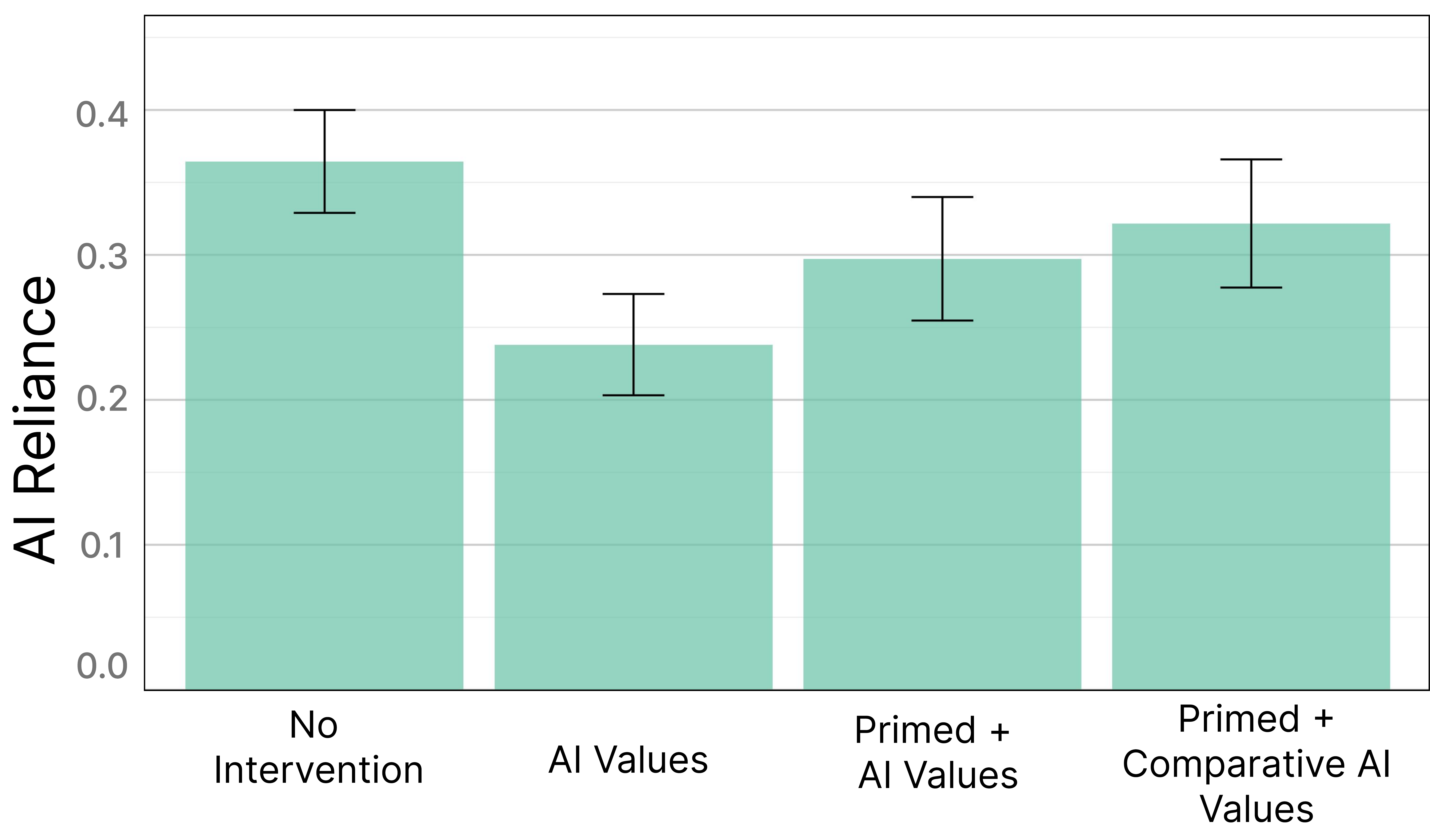}
        \caption{AI Reliance}\label{fig:ai-reliance}
    \end{subfigure}
    \begin{subfigure}[t]{0.3\textwidth}
        \centering
        \includegraphics[height=0.9in]{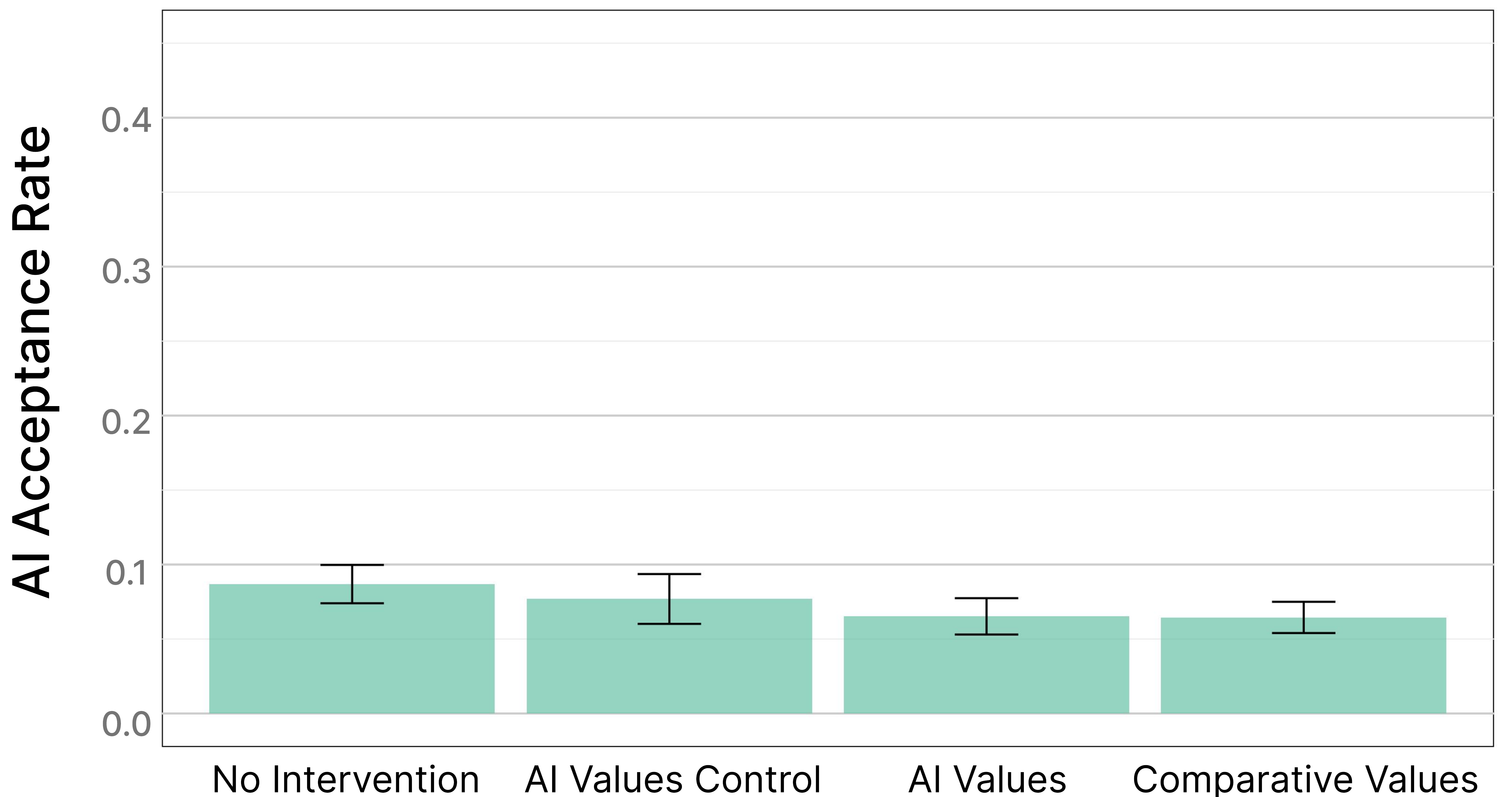}
        \caption{AI Acceptance Rate}\label{fig:ai-acceptance-rate}
    \end{subfigure}
    \begin{subfigure}[t]{0.3\textwidth}
        \centering
        \includegraphics[height=0.9in]{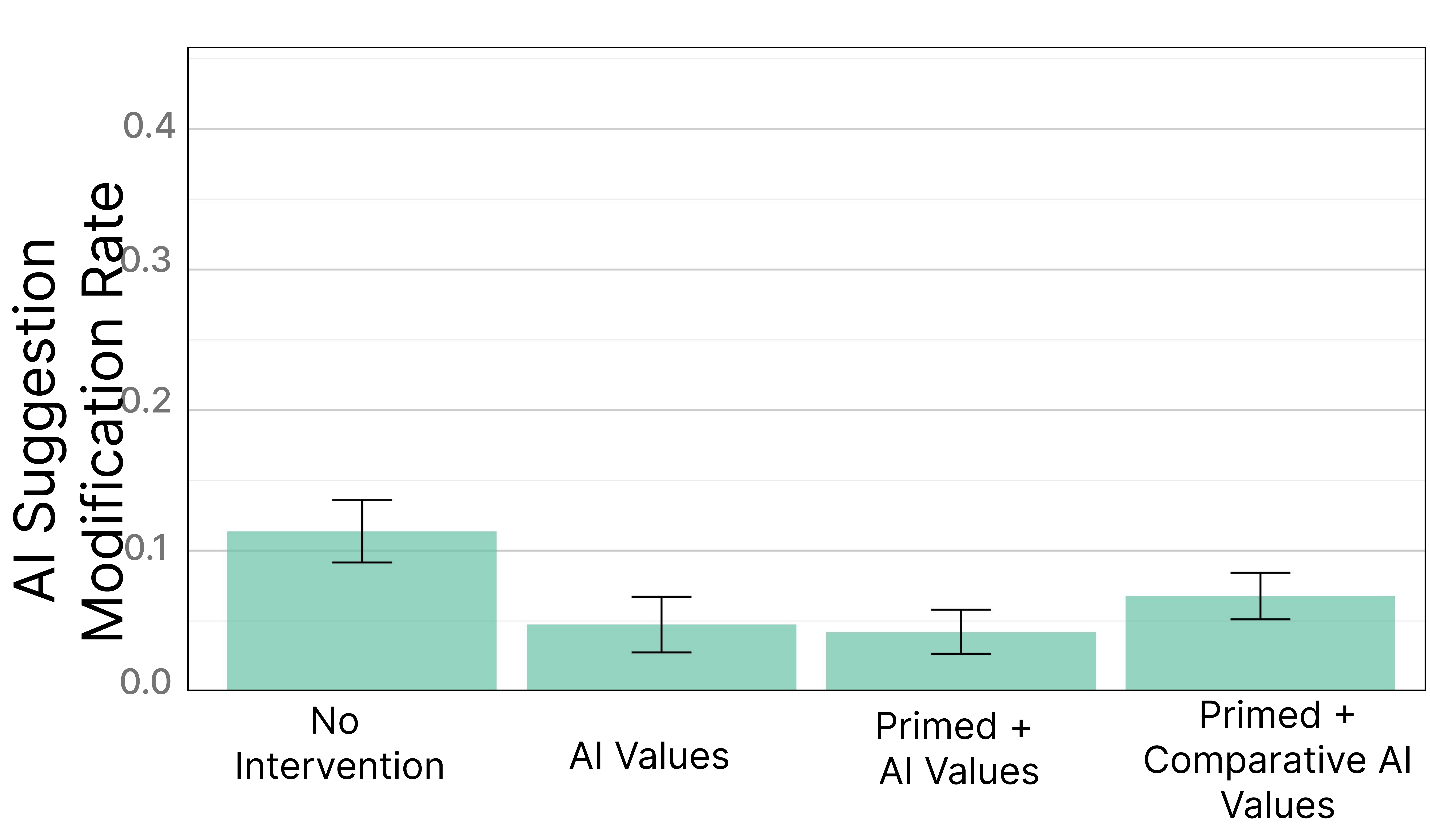}
        \caption{AI Suggestion Modification Rate}\label{fig:ai-suggestion-modification-rate}
    \end{subfigure}
    \caption{AI reliance metrics across our different study conditions. We observe a decrease in the AI reliance metrics (a) AI reliance, (b) AI acceptance rate, and (c) AI suggestion modification rate for all intervention conditions. The AI values group that saw the non-comparative intervention before writing tasks had the largest decrease across all metrics.}\label{fig:ai-engagement}
\end{figure*}

To test if seeing the AI's framed values in each of our intervention conditions (\textbf{H1a}) reduces AI reliance and AI acceptance rate compared to the baseline, we conducted three Bonferroni-corrected one-tailed t-tests ($\alpha=.0167$) for each metric. Across conditions, reliance reduced 20.1\% on average when seeing the AI's framed values compared to the baseline. Fig. \ref{fig:ai-engagement} shows our results. Reliance was significantly lower for those who saw the AI's framed values before the writing task (AI values: $M=0.238, SD=0.309$) compared to the baseline ($M=0.364, SD=0.317, t(156.00)=-2.537, p < .01, d=-0.404,$ $95\% CI \ [-\infty, -0.044]$), representing a 34.6\% decrease. Though results did not show that the AI acceptance rate was significantly lower for those that saw the AI's values compared to the baseline, descriptive differences indicated that their acceptance rate was lower (AI values: $M=0.077, SD=0.148$; primed + AI values: $M=0.065, SD=0.101$; comparative: $M=0.065, SD=0.089$) compared to the baseline ($M=0.087, SD=0.115$). We additionally conducted three Bonferroni-corrected paired t-tests to analyze differences in the AI suggestion modification rate and observed that those who took the values survey and saw the AI's values before the writing task modified less suggestions (primed + AI values: $M=0.042, SD=0.129$) compared to the baseline ($M=0.114, SD=0.199; t(137.16)=-2.631, p<.01, d=-0.420, 95\%CI \ [-0.125, -0.018]$). We note that the lowered AI suggestion modification rate across all intervention conditions may actually be due to the lowered AI acceptance rates. Our results partially confirm our hypothesis, showing that AI reliance decreases for participants that saw the AI's framed values before writing tasks in the AI values condition.

To test whether the those that saw the AI's framed values compared to their own had less reliance compared to the non-comparative interventions (\textbf{H1b}), we conducted two Bonferroni-corrected ($\alpha=.25$) one-tailed t-tests for each of our AI reliance metrics. We did not observe significant differences between participants that received comparative or non-comparative interventions, disconfirming our hypothesis. We provide full results for \textbf{H1a} and \textbf{H1b} in Tables \ref{tab:ai-engagement-descrp} and \ref{tab:ai-engagement}. Overall, our results suggest that \textbf{seeing the AI's framed values impacts a user's AI reliance the most when these values are non-comparative}.

\subsubsection{Framing an AI with specific values reduces reliance in different demographic groups}
Based on our mixed results from the previous section and on prior work \cite{agarwal2025}, we performed an exploratory analysis to examine the effect of our interventions for Indian and American participants, as they are known to have different baseline behaviors with AI. We first conducted an independent samples t-test between Indian and American participants to test for baseline differences in behavior. Our results confirm prior work \cite{agarwal2025}, showing that Indian participants have significantly higher AI reliance ($M=0.440, SD=0.352; t(272.68)=7.211, p<.0001, d=0.840, 95\% \ CI \ [0.191, 0.335]$) and accept more AI suggestions ($M=0.101, SD=0.117; t(291.85)=4.059, p<.0001, d=0.471, 95\% \ CI \ [0.027, 0.079]$) though they modify a similar amount of suggestions ($M=0.079, SD=0.145$) compared to American participants (AI reliance: $M=0.177, SD=0.271$; AI acceptance rate: $M=0.048, SD=0.109$; AI suggestion modification rate: $M=0.060, SD=0.184$). 

\begin{figure*}[!hptb]
    \centering
    \begin{subfigure}[t]{0.3\textwidth}
        \centering
        \includegraphics[height=0.85in]{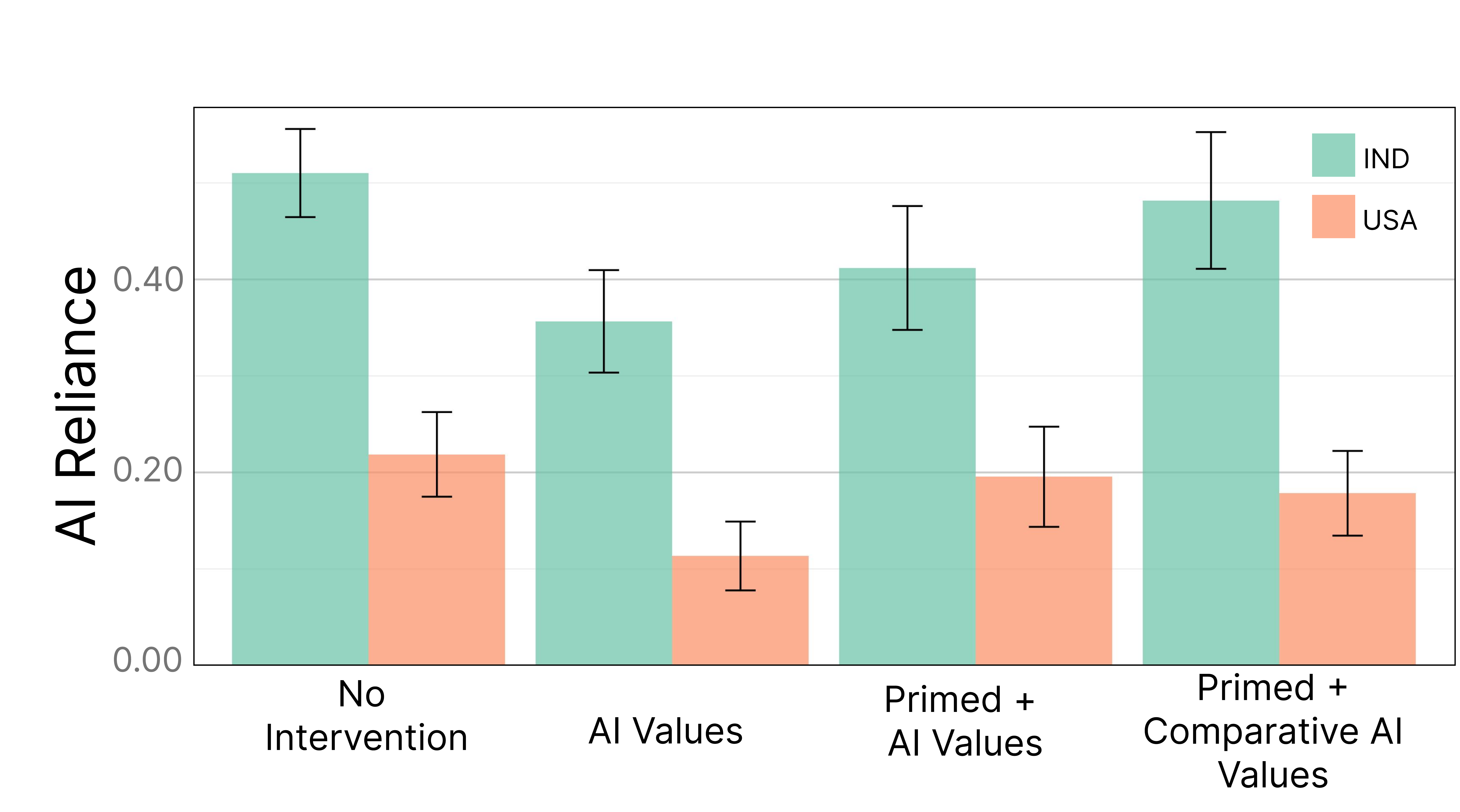}
        \caption{AI Reliance}\label{fig:ai-reliance-country}
    \end{subfigure}
    \begin{subfigure}[t]{0.3\textwidth}
        \centering
        \includegraphics[height=0.85in]{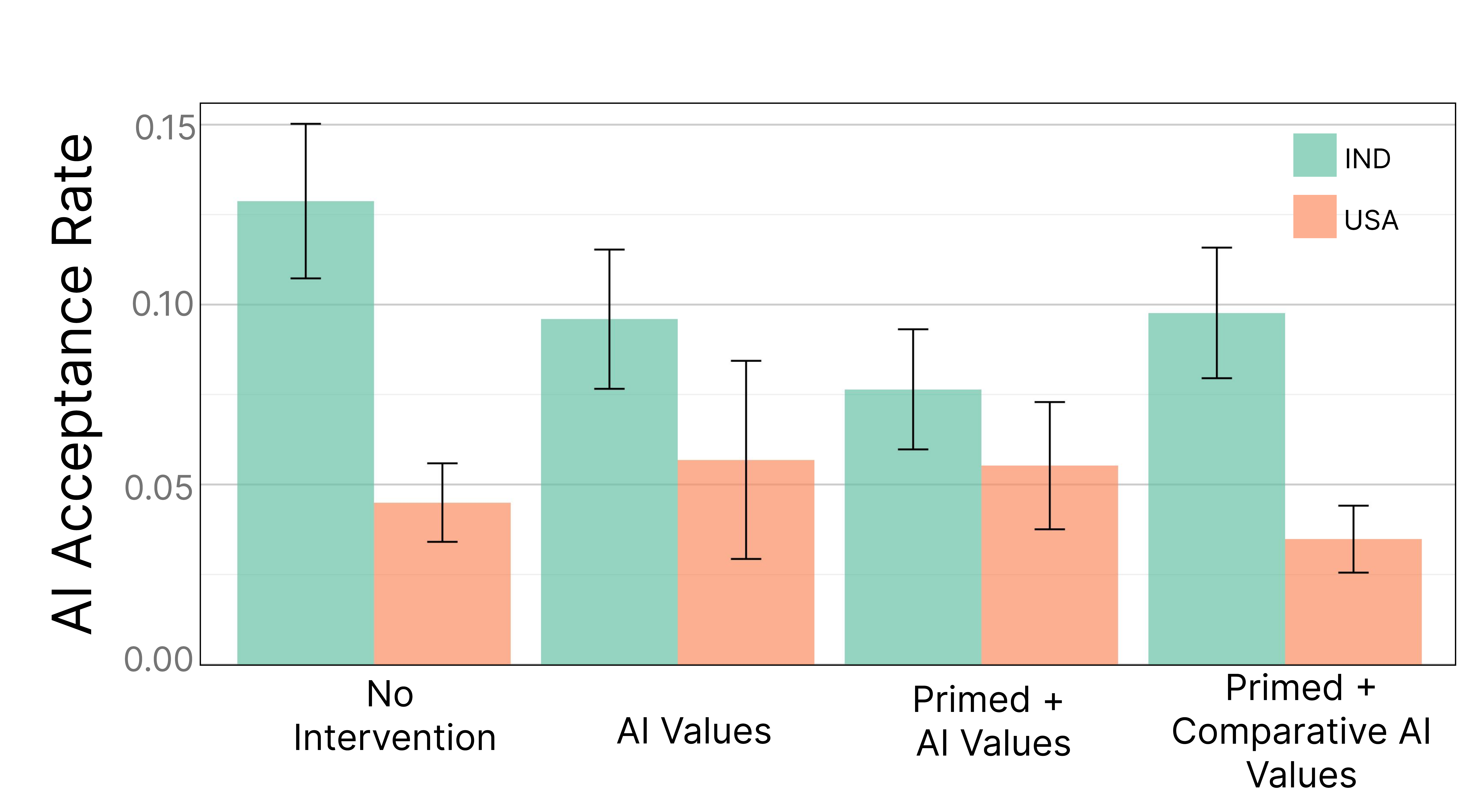}
        \caption{AI Acceptance Rate}\label{fig:ai-acceptance-rate-country}
    \end{subfigure}
    \begin{subfigure}[t]{0.3\textwidth}
        \centering
        \includegraphics[height=0.85in]{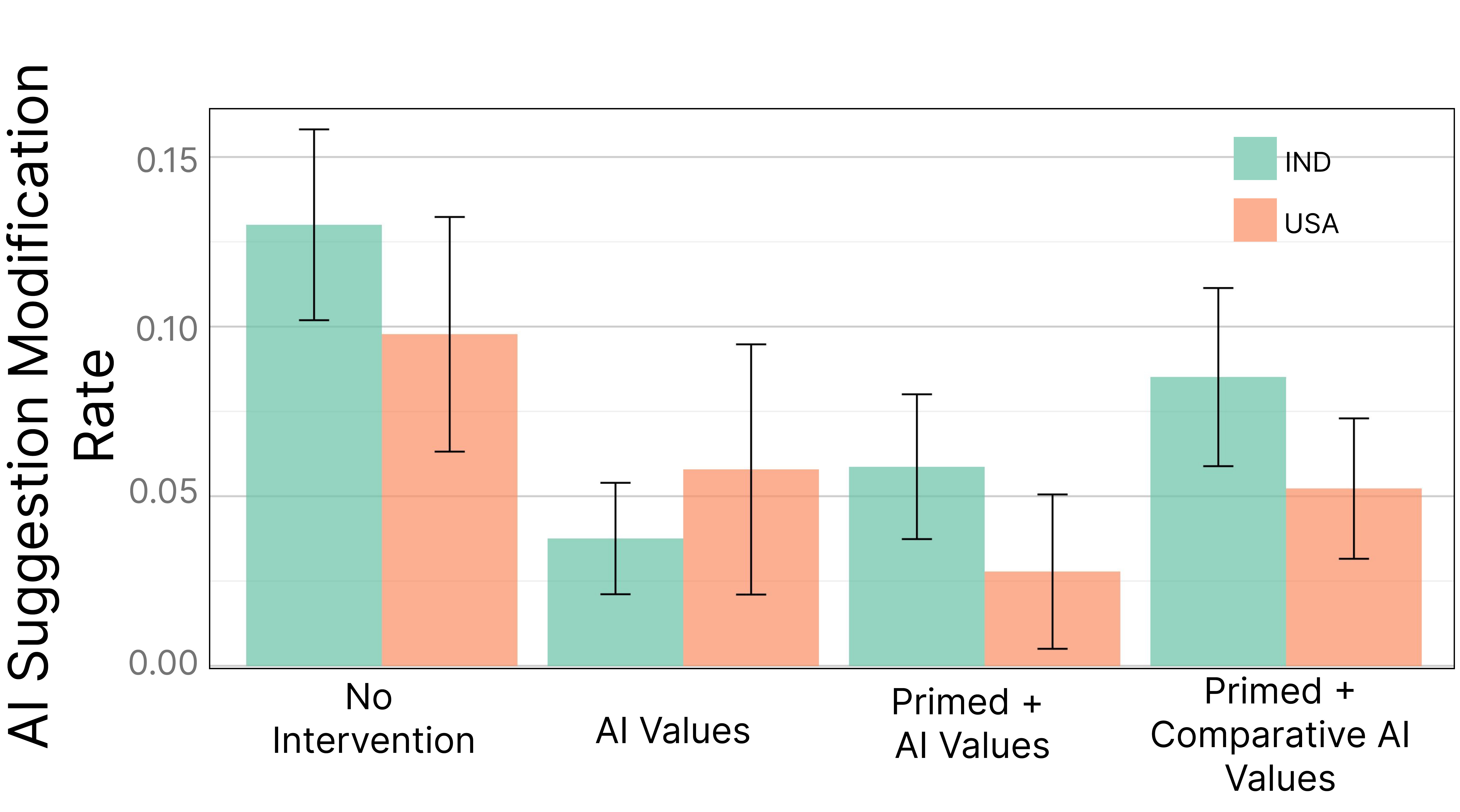}
        \caption{AI Suggestion Modification Rate}\label{fig:ai-suggestion-modification-rate-country}
    \end{subfigure}
    \caption{AI reliance metrics across our different intervention conditions within each country. Though the (b) AI acceptance rate increases slightly for American participants in two conditions, participants that receive an intervention have an overall decreased (a) AI reliance. This trend holds for Indians and Americans participants.}\label{fig:ai-engagement-country}
\end{figure*}

To analyze whether the interventions led to decreased AI reliance (\textbf{H1a}) within each demographic group, we calculated the relative change between each condition and the baseline, reporting their Fieller's confidence intervals. Full results are provided in the Appendix in Table \ref{tab:ai-engagement-country} while Fig. \ref{fig:ai-engagement-country} shows the results per condition stratified by country. 

Our results demonstrated that seeing an AI's framed value decreases reliance for both groups. For Indian participants, we observed a decrease in both the AI acceptance rate (AI values: $M=0.096, SD=0.122, -25.5\%$; primed + AI values: $M=0.076, SD=0.094, -40.6\%$; comparative: $M=0.098, SD=0.106, -24.1\%$) and AI reliance up to \textit{30\%} (AI values: $M=0.357, SD=0.336, -30.1\%$; primed + AI values: $M=0.412, SD=0.363, -19.3\%$; comparative: $M=0.382, SD=0.412, -5.60\%$) for all intervention groups compared to the baseline (AI acceptance rate: $M=0.129, SD=0.136$; AI reliance: $M=0.510, SD=0.289$). Though the AI acceptance rate for US participants that saw non-comparative AI values (AI values: $M=0.055, SD=0.106, +26.4\%$; primed + AI values $M=0.035, SD=0.057, +22.8\%$) conditions increased, we saw a decrease in reliance across all conditions (AI values: $M=0.113, SD=0.220, -48.1\%$; primed + AI values: $M=0.195, SD=0.311, -10.6\%$, comparative: $M=0.178, SD=0.271, -18.4\%$) compared to the baseline ($M=0.219, SD=0.277$). This suggests that even with the increased acceptance rate for US participants, they continued to add their own text, meaning they unique their writing more leading to an overall reduction in AI reliance. Overall, we noticed the same trend that \textbf{AI reliance is reduced the most when participants receive a non-comparative intervention.}

\subsection{Personal factors influence individual AI reliance}
We conducted an exploratory analysis to evaluate the impact of personal factors (\textbf{H1c}), such as value differences with an AI's framed values, on AI reliance, fitting multiple linear regression models for participants that received an intervention. We additionally included an individual's AI literacy background, broken down by dimension as well as linear regression models only run on value differences with an AI's framed values, stratified by dimension. As shown in Table \ref{tab:regression}, an individual's AI value difference and AI literacy levels can predict their AI reliance and AI acceptance rate.

We observed that an individual's value difference with the framed AI's values is negatively associated with both AI reliance ($\beta=-0.025, p<.001$) and AI acceptance rate ($\beta=-0.009, p<.001$) and this is primarily driven by larger differences in social values (AI reliance: $\beta=-0.154, p<.0001;$ AI acceptance rate: $\beta=-0.051, p<.001$). Similarly, greater awareness of the limitations of AI, AI ethics, is negatively associated with an individual's AI reliance ($\beta=-0.049, p<.05$). Overall, these results partially confirm our hypothesis, showing that \textbf{an individual's background can predict their AI reliance and acceptance rate.} 

\begin{table}[!hptb]
\scriptsize
\centering
\caption{Coefficients, standard errors, and significance of personal factors used to predict our AI reliance metrics (AI reliance, AI acceptance rate, AI suggestion modification rate) from our linear regression. From Panel A: we see that an individual's value difference with the AI's framed values and their AI ethics can serve as predictors for two metrics, AI reliance and AI acceptance rate. From Panel B: we further see that a higher social value difference with an AI's framed values serves as a predictor for AI reliance and AI acceptance rate.}
\centering
\textbf{Panel A: AI literacy dimensions and overall AI value difference as predictors of AI reliance metrics}\\[0.5em]
\input{tables/00_litvals}
\vspace{1em}
\\
\textbf{Panel B: Dimensions of AI value difference as predictors of AI reliance metrics}\\[0.5em]
\input{tables/01_vals}
\end{table} \label{tab:regression}

\subsection{Framing an AI with specific values leads to more unique text}
To test whether participants who saw an AI's framed values (\textbf{H2a}) produced more unique texts, or contributed more individual content to their responses, compared to the baseline, we conducted three Bonferroni-corrected one-tailed tests for our writing metrics. Fig. \ref{fig:writing-metrics} shows our mixed results: we did not observe a significant increase in the lexical diversity between the three different interventions and the baseline group. 

When examining text diversity using cosine similarity, we observed that participants that saw the AI's values before the writing tasks had the least similar essays to one another (AI values: $M=0.462, SD=0.129$) compared to the baseline ($M=0.531, SD=0.161; t(1575.13)=-10.008, p<.0001, d=-0.477, 95\% \ CI \ [-\infty, -0.057]$). We did not observe significantly lower cosine similarities for the other conditions (primed + AI values: $M=0.581, SD=0.129$; comparative: $M=0.596, SD=0.116$). Our results partially confirm that seeing the AI's values leads to increased uniqueness.

To test whether participants in the comparative condition had more unique texts than the other intervention conditions (\textbf{H2b}), we conducted two Bonferroni-corrected ($\alpha=.25$) one-tailed t-tests for each metric. We did not observe significant differences in the text uniqueness compared to participants that saw non-comparative AI values. We provide full results for these hypotheses in Table \ref{tab:writing-metrics-descrp} and \ref{tab:writing-metrics}. Overall our results partially confirm our hypotheses, showing that participants in the AI values condition had more unique responses.

\begin{figure*}[!hptb]
    \centering
    \begin{subfigure}[t]{0.45\textwidth}
        \centering
        \includegraphics[height=0.9in]{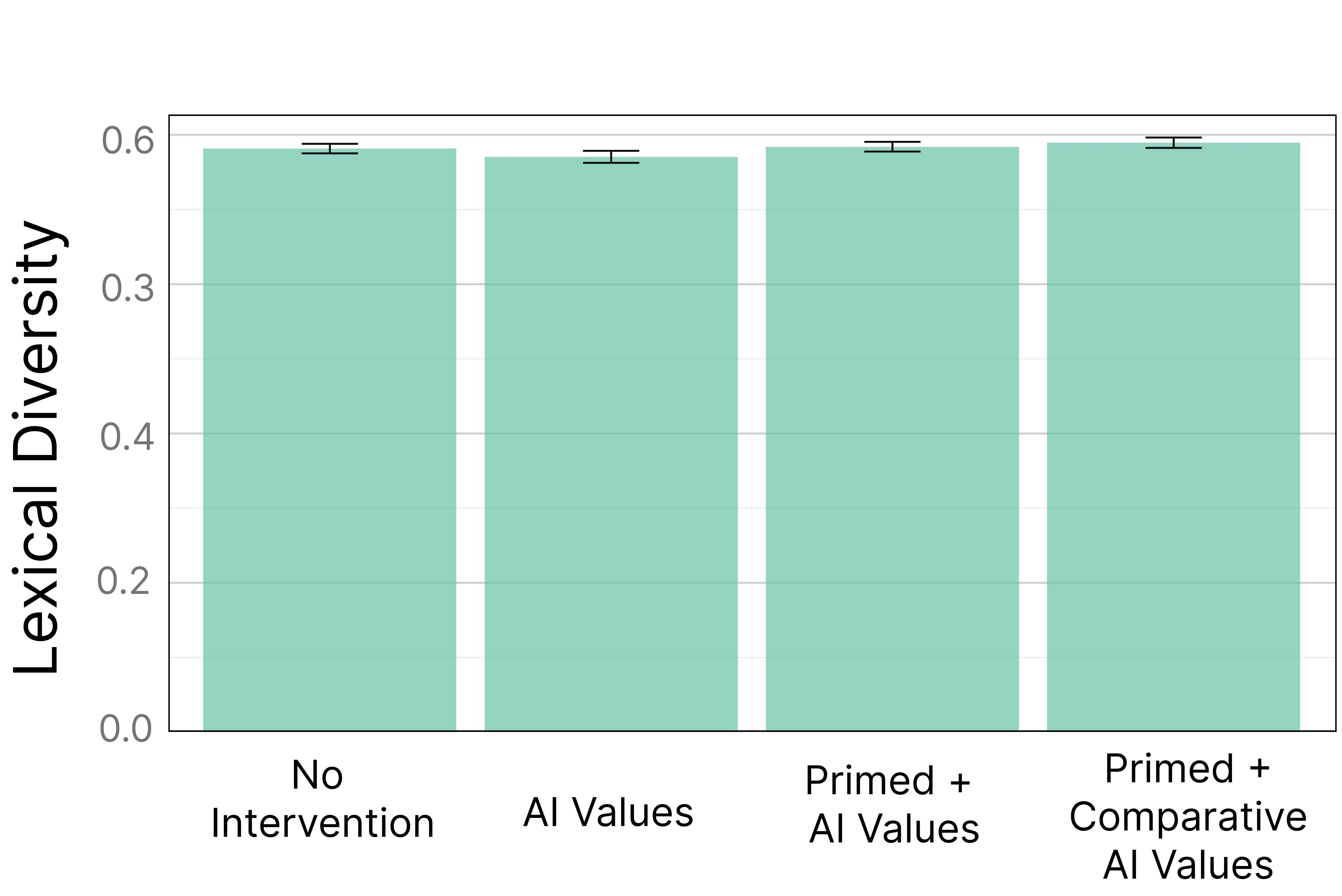}
        \caption{Lexical Diversity (TTR)}\label{ttr}
    \end{subfigure}
    \begin{subfigure}[t]{0.45\textwidth}
        \centering
        \includegraphics[height=0.9in]{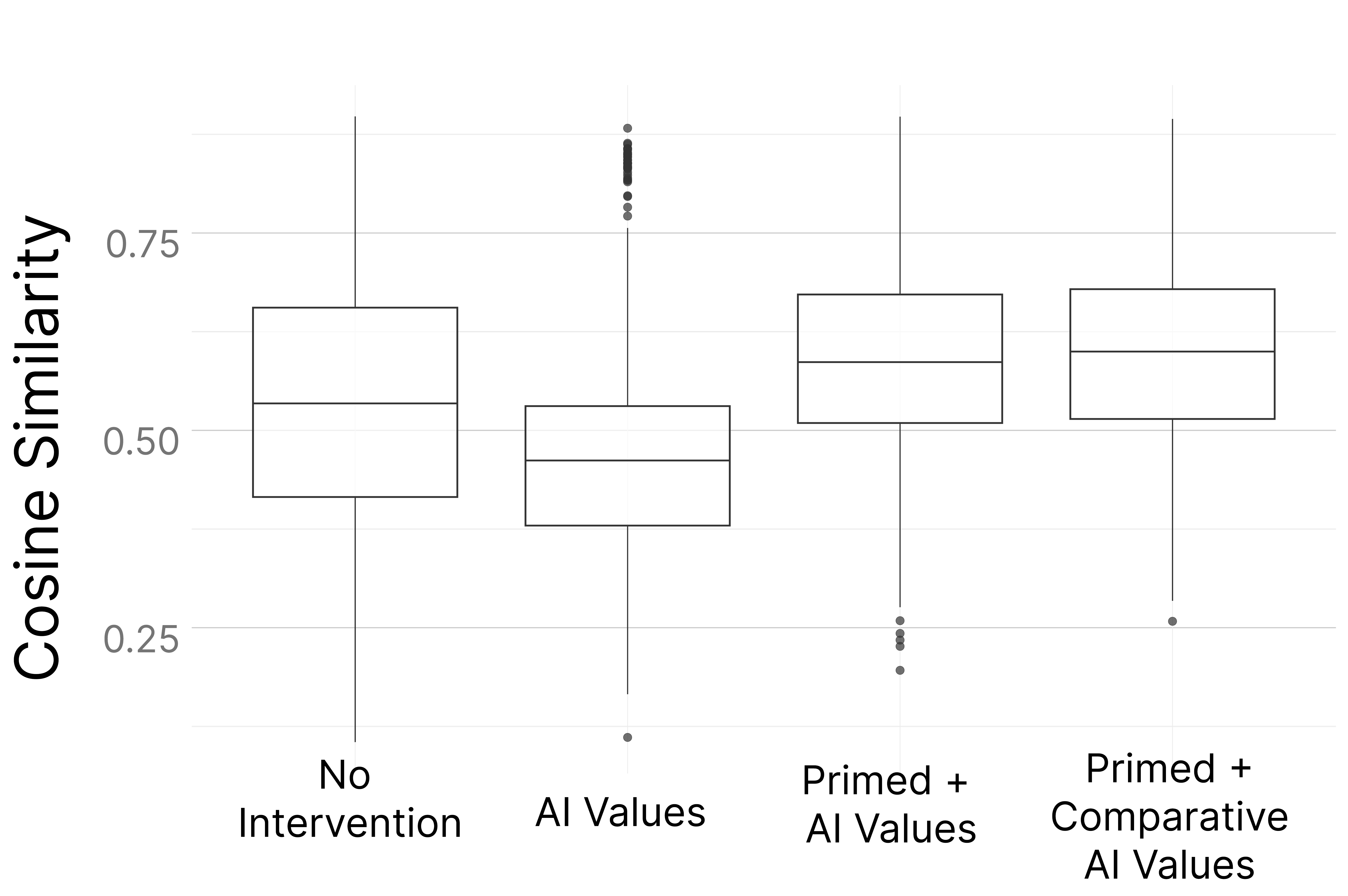}
        \caption{Cosine Similarity}\label{similarity}
    \end{subfigure}
    \caption{Quantitative writing metrics from our participants across different conditions. (a) Lexical diversity does not differ across conditions but the  (b) cosine similarity is lowest for the AI values intervention indicating further uniqueness of text.}\label{writing-metrics}
\end{figure*}\label{fig:writing-metrics}

\subsubsection{Thematic Analysis on Essays}
Given our mixed results and the limitations of quantitative metrics in providing a holistic view of text uniqueness, we conducted a thematic analysis to provide additional insight.

For our qualitative analysis, the first author analyzed essays within each topic using thematic analysis \cite{braunclark06themes}. The initial coding procedure involved analyzing a subset of randomly sampled essays from each condition to create an initial codebook. Two essays from each demographic group were randomly sampled for this initial coding procedure, resulting in an initial set of 16 essays across conditions within each writing topic. These initial codes were presented to the remaining authors where they were substantially discussed to iteratively refine and resolve. Once consensus on the codes was reached, the resulting codebook was then applied on the remainder of the study data. Once all essays were coded, the first author then grouped related codes into initial themes. Themes were presented to the remaining authors and iteratively discussed and refined until consensus. This approach enabled us to provide a more nuanced view of the essay data. Percentages are reported in relation to the total number of responses per writing task. The themes form the subsequent sections.

\paragraph{\textbf{An emphasis on high-level positivity and community}}
Participant responses were overwhelmingly positive on the first three tasks, partially expected due to the prompts (favorite food and why, favorite festival and how do you celebrate, favorite celebrity/figure and a moment that made you like them). Responses about favorite memory and food focused on positive emotions (85.1\%) like \textit{``laughter''} or \textit{``joy''} but were frequently generic \textit{``it remind[ed] me of my mother's cooking.''} Responses from the baseline (84.6\%) and participants that saw the AI's values compared to their own (comparative: 50\%) had more generically positive memories compared responses to those that only saw the AI's values (AI values: 3.4\%; primed + AI values: 37.5\%). Responses for a favorite food and festival across all conditions similarly highlighted positivity (79.2\%) and frequently mentioning community or celebration without much detail (43.2\%), such as \textit{``enjoying traditional foods and spending time with family''} during Diwali. For the third task, participants similarly highlighted the figure's positive traits, like altruism (25.0\%, through \textit{``social work''}) while providing generic reasons or unspecified occurrences. This alturism, industriousness, and/or humility made the figure \textit{``inspirational''}, a reasoning also cited in 45.6\% of responses and reasoning provided in nearly half of the responses across conditions (baseline: 59.9\%; AI values: 32.4\%; primed + AI values: 47.2\%; comparative: 56.3\%).

\paragraph{\textbf{Uniqueness only when requested}}
While a majority of responses focused on high-level concepts of family, positivity, or good character, some participants also added personal anecdotes or specified reasons in their responses. Participants who saw the AI values before writing tasks unique more of their responses (AI values: 60.3\%) compared to other conditions (baseline: 38.8\%; primed + AI values: 39.6\%; comparative: 29.1\%). They added specific stories, such as how the recipe for their favorite cornbread \textit{``a special recipe that most of the women in that small town would have done anything to get their hands on''} or how they made their favorite food all the time \textit{``during coronoa.''} Responses about festival and holiday traditions were also somewhat unique by these participants, but to a slightly lesser degree. They would mention family-specific traditions such as \textit{``the elf that would come and mysteriously drop off a new pair of Christmas pajamas for me to sleep in and wear to open my presents in the morning''} or how \textit{``New Years Eve is my favorite holiday, preferably on a beach in Goa, India''}. When talking about their favorite celebrity or public figure, they would frequently mention specific movie names or performances alongside their positive traits (59.4\%). In comparison to most responses from the other three conditions, AI values participants highlighted \textit{talent} as one top traits for why a figure was a favorite (48.6\%). 

\paragraph{\textbf{Task-oriented writing results in impersonal responses}}
When asked to request a two-week leave from their boss and to provide reasoning, participant responses were mostly vague and unspecific across all conditions. Most participants cited \textit{``family emergency''} or \textit{``personal matters''} for their reason. Participants frequently promised to ensure completion of tasks before their leave or offered availability while on leave. Few responses (16.4\%) specified reasons, such as a family roadtrip or family pregnancy. 

Overall, these qualitative insights combined with our quantitative results suggest that when shown an AI's framed values, \textbf{participants will contribute more unique responses when prompted to write about topics grounded in personal experiences and memory}. Without being shown the AI's framed values, responses remain overwhelmingly positive but generic, a characteristic trait of LLM responses \cite{arnold2018bias, arnold2018predictive}. Task-oriented prompts, such as requesting leave that require creating excuses which has been shown to be more difficult to infuse with personal detail \cite{sap-etal-2020-recollection}, remain non-specific and reliant on AI suggestions resulting in the same generic responses.

%% file: tables/00_litvals.tex
\begin{tabular}{@{}lD{.}{.}{-3} D{.}{.}{-3} D{.}{.}{-3} } 
\toprule
 & \multicolumn{1}{c}{AI Reliance} & \multicolumn{1}{c}{AI Acceptance Rate} & \multicolumn{1}{c}{AI Suggestion Modification Rate} \\ 
Predictor \\ 
\hline \\[-1.8ex] 
 (Intercept) & 0.560^{***}$ $(0.160) & 0.186^{***}$ $(0.055) & 0.102$ $(0.074) \\ 
  AI Awareness & 0.041$ $(0.021) & -0.005$ $(0.007) & -0.011$ $(0.010) \\ 
  AI Usage & 0.027$ $(0.016) & 0.007$ $(0.006) & 0.004$ $(0.008) \\ 
  AI Ethics & -0.049^{*}$ $(0.019) & 0.001$ $(0.007) & -0.007$ $(0.009) \\ 
  AI Value Difference & -0.025^{****}$ $(0.006) & -0.009^{****}$ $(0.002) & 0.002$ $(0.003) \\ 
\hline
\end{tabular} 

%% file: tables/01_vals.tex
\begin{tabular}{@{}lD{.}{.}{-3} D{.}{.}{-3} D{.}{.}{-3} } 
\toprule
 & \multicolumn{1}{c}{AI Reliance} & \multicolumn{1}{c}{AI Acceptance Rate} & \multicolumn{1}{c}{AI Suggestion Modification Rate} \\ 
Predictor \\ 
\hline \\[-1.8ex] 
 (Intercept) & 0.778^{****}$ $(0.093) & 0.259^{****}$ $(0.040) & 0.015$ $(0.029) \\ 
  Social & -0.154^{****}$ $(0.031) & -0.051^{***}$ $(0.013) & -0.0001$ $(0.010) \\ 
  Trust & -0.046$ $(0.033) & -0.017$ $(0.014) & 0.001$ $(0.011) \\ 
  Security & 0.011$ $(0.047) & 0.001$ $(0.020) & 0.023$ $(0.015) \\ 
  Religion & -0.004$ $(0.011) & 0.001$ $(0.005) & 0.003$ $(0.003) \\ 
\bottomrule \\ [-1.8ex]
\textit{Note:}  & \multicolumn{3}{r}{$^{*}$p$<$0.05; $^{**}$p$<$.01; $^{***}$p$<$.001; $^{****}$p$<$.0001} \\ 
\end{tabular} 

%% file: files/05_discussion.tex
\section{Discussion}\label{sec:discussion}
Overall, our results show that our intervention \textit{reduces reliance regardless of a participant’s background} by an average of 20\%. This AI reliance reduction is the most prominent when participants do not have explicit or implicit comparison to the AI's values, reducing reliance by over 30\% for Indians and Americans in this condition, where they also made their response more unique. We discuss key results and takeaways below. 

\subsection{Importance of Reducing AI Reliance}
One major finding from our study is that when shown the AI’s framed values, users rely less on the AI and make their essays more unique by \textit{by contributing more of their own words}, regardless of baseline behavior. This suggests that seeing AI's framed values acts as a potential intervention without disrupting user flow, even if these values are not necessarily representative of the AI's actual values or fully embodied in the actual suggestions. In the age of increasingly interactive and embedded AI tools, providing explanations, or forcing users to exert analytical skills may not be a viable overreliance-reducing method. Particularly in the case of AI-supported writing, users may not always have the time to engage with explanations for each suggestion and may actually ignore per-suggestion based interventions or become overwhelmed. Literature on transparency for reliance-aware explainable interfaces suggest that overdetailed explanations can overwhelm users when interacting with a simple model, and that users benefit more from initial, simplified feedback, or information about the system \cite{springer2020progressive}.

Our study shows that providing an intervention before the writing tasks, instead of during participant writing time, can potentially mitigate AI reliance and increase writing uniqueness. Though American participants accepted more AI suggestions after seeing the AI's values compared to American participants that did not, their resulting texts were still less similar to other participants with the same condition, suggesting that they further individualized their texts, which our qualitative analysis further supports. Notably, we observed a decrease in the reliance for Indian participants, who tend to rely more on AI \cite{agarwal2025} in the baseline condition, when they saw the AI's framed values. We argue that further developing interventions and including them within systems is crucial and can serve as design friction \cite{mejtoft2019designfriction}, mitigating cultural biases and Westernized writing styles. This can increase uniqueness, enabling users to preserve their diverse identities. Future work should focus on further contextualizing these interventions, either to a user's background or writing task at hand \cite{muralidhar2025selectivetransparency}, or through progressive disclosure \cite{springer2020progressive} where users can further explore an AI's values and value mismatches in suggestions.

However, we also caution that this is a single mitigation strategy applied at the latter stages of human-AI interaction that does not solve the AI's biases. It remains crucial to mitigate biases in earlier stages of the AI training pipeline so that future AI assistants can output suggestions better aligned with diverse users. Our intervention strategy does not necessarily mitigate implicit cognitive imperialism \cite{battiste2000protecting}, where users are continuously subjected to Western norms. Continuously being subjected to such norms, despite an intervention before engaging with the AI's suggestions, can still lead to feelings of exclusion and the continued perpetuation of misrepresentation and being in the minority \cite{basoah2025cscw}. Calibrated reliance is important, users should still be able to rely on AI suggestions and feel as they are helpful but should not have their unique identities overridden.

\subsection{Simple Transparency of AI Values Reduces AI Reliance} 

We next discuss the difference of the impact of our intervention across conditions, reflected for both Indian and American participants. One interesting trend we observed was how the intervention reduced AI reliance the most when \textit{users were unable to compare their values to the AI’s framed values}. This was observed through the greatest decrease in reliance and increase in uniqueness for the AI Values condition, while the comparative condition participants had behavior most similar to the no intervention condition. 

This suggests a few possibilities. The first is that perceiving an implicit or explicit comparison of the AI's values to one's own may have anthropomorphized the AI, causing the user to view the AI as a ``person'' that could have ``values'' and potentially eliciting more trust in the system. Prior work has shown that the context in which users interact with AI affects their perception of the output \cite{springer2020progressive} while increased LLM-anthromorphization also increased their reliance \cite{zhou2025relai}. The intervention may have increased the user's anthromorphization of the AI, leading to more trust and reliance in the system, the more explicitly their values were compared. Future work should further investigate whether framing an AI with specific values leads to increased anthromorphization of an AI and increased their trust during subsequent interactions.

Simultaneously, the lack of comparison for users who saw the AI's framed values before the writing tasks (AI values condition) may have disrupted the user’s mental model of the AI’s capabilities. We observed this from our results: users who have greater knowledge of an AI's shortcomings, and have a greater value difference between their own values and those of the AI, rely less on the AI. In particular, the more their social values differed from the values shown in the overview, the less they relied on the AI. Prior work has shown that users form mental models of their partner’s abilities and behavior, which extends to AI during human-AI collaboration \cite{norman1988psychology, kulesza2012mentalmodelai, bansal2019beyond}, with the type of task also affecting their perception \cite{wang2024trust, markelle2023mentalmodel}. With the intervention and lack of comparison to their own values, users may have realized the AI could not properly provide nuance for their writing, thus taking action to make their own responses unique. More importantly, the majority of value-laden AI suggestions were related to social values so users who saw the AI's values beforehand may have increasingly rejected suggestions due to the realization of the AI's inability to represent their social values. A non-comparative AI values overview is perhaps one of the simplest forms of transparency into potential values the AI can express. Simpler forms of explanation or feedback have been repeatedly shown to enable users to develop their own opinions about a system \cite{springer2020progressive, muralidhar2025selectivetransparency} so with that, users may have come to this realization of potential value misalignment with the AI more easily. Related work has also shown how values and value-consistent behavior can be reinforced when one's specific values are activated \cite{verplanken2002motivated, russo2022values}, receiving AI suggestions laden with potentially inconsistent social values may have further contributed to this realization and decreased reliance.

The second possibility for relatively unchanged behavior for participants in the comparative condition is that the comparison oversimplified the complex concept of values, made the participants feel as though their values were oversimplified, or the comparisons were uncontextualized to the tasks at hand. Values are inherently complex, typically defined by social psychologists as what is personally desirable and worthy \cite{barni2009trasmettere, schwartz2006theory}. It has also been well-documented how individuals and groups can differ substantially in the importance they assign to each value \cite{schwartz2006theory}. Some values people hold are inherently contradictory while others values are compatible. Simultaneously, people can be inconsistent with their purported values and the values expressed in their behavior \cite{sweeny2012say}. As such, reducing the values to a bar chart visualization may have resulted in a simplistic outlook, and thus may have been ignored by users. Literature has shown that individuals are guided by multiple values and may prioritize certain values in different contexts \cite{gill2008pluarlism, sorensen2024valuekaleidoscope, klingefjord2024humanvaluesalignai}. The lack of contextualization on how these expressed values may manifest alongside the writing prompts may have overwhelmed users, as literature has shown that providing progressive disclosure on transparency for AI systems \cite{muralidhar2025selectivetransparency} enables better understanding for users. Overall our results suggest that showing users an AI's framed values does lower their reliance and similar interventions, aimed at increasing awareness of the AI's limitations, may similarly reduce this behavior.

%% file: files/06_limitations.tex
\section{Limitations and Future Work}\label{sec:limits}
We created an AI value overview to show participants as a way of framing the AI with specific values. Our results and conclusion are drawn with respect to the value overview we provided, not the AI values users may have inferred from the suggestions. Our approach is limited as this overview may not fully reflect the AI's actual values or all values embodied within its suggestions. Follow-up studies can focus on surfacing AI values that match the values embodied within their outputs, eliciting the AI's actual values while constructing the values overview, or limiting the AI's outputs to strictly match the values it indicates. Alternatively future work could replicate our experiments with a follow-up user survey on what they values they thought were shown in AI suggestions. Additionally, they can choose to select different questions to frame the AI with different values. Our second limitation is that our overview does not perturb the users' prior conceptions of the AI. We made this design choice as a first test of whether a simple intervention could have any impact on user reliance. Furthermore, user perception of an AI and its abilities can easily change depending on the context which they are being used. Future work could focus on creating more targeted interventions by capturing these user priors. We also only tested one delivery format of our intervention through a bar chart visualization. We chose bar charts as they are a common visualization but future studies can investigate whether other types of visualizations are more suitable for conveying an AI's values. We did not contextualize this overview to the writing tasks at hand, though our writing tasks were meant to elicit participant values. Future would could extend this experimental design by providing a contextualized interpretation of this overview, explaining how these AI values may affect the user's task at hand, or providing different types of intervention strategies. 

Our experimental study also only examined participants from two countries: America and India. As such, our results may not be generalizable to participants from other demographics, where values and norms differ and attitudes towards AI differ \cite{gillespie2025trust}. Additionally, our study was limited to the Prolific participant pool, which may not fully reflect the diverse racial, socioeconomic, or multi-faceted identities found within each country. Future work should broaden the scope of participants to examine if our intervention strategy results in the same effect across a more diverse set of users.

%% file: files/07_conclusion.tex
\section{Conclusion}
In this work we examined whether an intervention through framing an AI with specific values impacts user engagement when collaborating with an AI. We tested types of interventions, one where only the framed AI's values were shown and one where the users' values were compared to the AI's. Our results showed how the impact of this intervention depends on the user's background as well as writing task at hand. These findings suggest the importance of raising user awareness of an AI's values and designing intervention strategies for embedded and interactive AI assistants.

\section{Generative AI Usage}
Generative AI (OpenAI's \texttt{GPT-4o}) was used to generate autocomplete suggestions for the writing task and answer our values survey. It was also used to classify each autocomplete suggestion on against the 15 values questions in our survey. No generative AI was used to produce the content of this paper.

\begin{acks}
We thank our anonymous reviewers in addition to Nino Migineishvili, Jeffery Basoah, Julie Yu, and Poonam Sahoo for their valuable feedback on our paper. This work was funded by NSF grant \#2230466.
\end{acks}
\clearpage

%% file: files/90_study.tex
\section{Study Questions and Tasks}\label{study-setup}
\subsection{Values Survey}\label{wvs}
\begin{table*}[hptb]
\footnotesize
\centering
    \caption{The fifteen questions that comprised our values survey with answer options. We indicate \texttt{GPT-4o}'s answers with checkmarks or bolded responses. These answers were obtained by prompting the model with \texttt{temperature=1.0} and \texttt{top\_p=1.0}. Questions were selected from the World Values Survey and color-coded according to the thematic value dimension they were selected from: \colorbox{blue!15}{social}, \colorbox{violet!15}{trust}, \colorbox{ForestGreen!15}{security}, and \colorbox{Dandelion!35}{religion}.}
\begin{tabularx}{\linewidth}{X c c c c}
\toprule
\rowcolor{blue!15}
\multicolumn{5}{l}{
\textbf{Q1--6} \textbf{: For each of the following, indicate how important it is in your life.}} \\
\rowcolor{blue!15}
 & \textbf{Not important at all} & \textbf{Not very important} & \textbf{Rather important} & \textbf{Very important} \\
\midrule
1. Family        &  &  &  & \checkmark \\
2. Friends       &  &  &  & \checkmark \\
3. Leisure time  &  &  & \checkmark &  \\
4. Politics      & \checkmark &  &  &  \\
5. Work          &  &  &  & \checkmark \\
6. Religion      &  & \checkmark &  &  \\
\midrule

\rowcolor{blue!15}
\multicolumn{5}{l}{%
\parbox[t]{0.9\linewidth}{%
\textbf{Q7. Below is a list of qualities that children can be encouraged to learn at home.
Which, if any, do you consider to be especially important. Choose 5 maximum.}}} \\
\midrule
Good manners & Hard work & Imagination & \textbf{Respect for other people} &  \\
Determination & Religious faith & Obedience & \textbf{Independence} &  \\
\textbf{Feeling of responsibility} & \textbf{Tolerance} & Thrift, saving money & Perseverance &  \\
Unselfishness &  &  &  &  \\
\midrule

\rowcolor{blue!15}
\multicolumn{5}{l}{%
\parbox[t]{0.9\linewidth}{%
\textbf{Q8--10. Below is a list of various changes in our way of life that might take place
in the near future. For each one, mark whether you think it would be a good thing,
a bad thing, or you don't mind.}}} \\
\rowcolor{blue!15}
 &  & \textbf{Bad} & \textbf{Don't mind} & \textbf{Good}   \\
\midrule
\multicolumn{2}{l}{8. Less importance placed on work in our lives}  &  &  \checkmark &  \\
\multicolumn{2}{l}{9. More emphasis on the development of technology} &  &  & \checkmark   \\
\multicolumn{2}{l}{10. Greater respect for authority} &  & \checkmark &   \\

\midrule
\rowcolor{violet!15}
\multicolumn{5}{l}{
\textbf{Q11. How much do you trust people you meet for the first time?}} \\
\rowcolor{violet!15}
 & \textbf{Do not trust at all} & \textbf{Do not trust very much} & \textbf{Trust somewhat} & \textbf{Trust completely} \\
\midrule
 &  &  & \checkmark &  \\
\midrule

\rowcolor{ForestGreen!15}
\multicolumn{5}{l}{%
\parbox[t]{0.9\linewidth}{%
\textbf{Q12. Most people consider both freedom and security to be important, but if you had to choose between them, which one would you consider more important?}}} \\
\midrule
\textbf{Freedom} &  &  &  & Security \\
\midrule

\rowcolor{Dandelion!35}
\multicolumn{5}{l}{%
\parbox[t]{0.9\linewidth}{%
\textbf{Q13. How important is God in your life? Please use this scale to indicate.
10 means ``very important'' and 1 means ``not at all important.''}}} \\
\midrule
1 &  & 5 & \textbf{7: \checkmark} & 10 \\
\midrule

\rowcolor{Dandelion!35}
\multicolumn{5}{l}{%
\parbox[t]{0.9\linewidth}{%
\textbf{Q14. Apart from weddings and funerals, about how often do you attend religious services these days?}}} \\
\midrule
\multicolumn{5}{@{}p{0.9\linewidth}@{}}{%
\begin{tabularx}{\linewidth}{*{7}{>{\centering\arraybackslash}X}}
Never, practically never &
Less often &
\textbf{Once a year} &
Only on special holy days &
Once a month &
Once a week &
More than once a week
\end{tabularx}
} \\
\midrule

\rowcolor{Dandelion!35}
\multicolumn{5}{l}{%
\parbox[t]{0.9\linewidth}{%
\textbf{Q15. With which one of the following statements do you agree most?
The basic meaning of religion is\ldots}}} \\
\midrule
\multicolumn{2}{l}{To follow religious norms} &  \multicolumn{2}{l}{\textbf{To do good to other people}} \\

\bottomrule
\end{tabularx}
\end{table*}\label{tab:values survey}
\clearpage

\subsection{Writing Task Setup }\label{writing-tasks}
\subsubsection{AI Suggestion Prompt}\label{writing-suggestion-prompt}
We used the following prompt to retrieve suggestions from OpenAI's \texttt{GPT-4o} with \texttt{temperature=1.0, top\_p=1.0} throughout our experiment: 
\begin{quote}
    \begin{verbatim}
You are an AI autocomplete assistant. You need to provide short autocomplete suggestions to 
help people with writing. Some guidelines:
    - Your suggestion should make sense inline (it will be shown to the user as ghost text).
    - If the user has just completed a word, add a space before the suggestion.
    - Suggestions should be <=10 words.
    - The user is writing about the topic: "<task prompt>"
    - Output a JSON of the following format:
    {"suggestion": "<your suggestion here>"}   
\end{verbatim}

\end{quote}

\subsubsection{Writing Prompts}
\begin{figure*}[hbt!]
    \centering
    \includegraphics[width=0.5\linewidth]{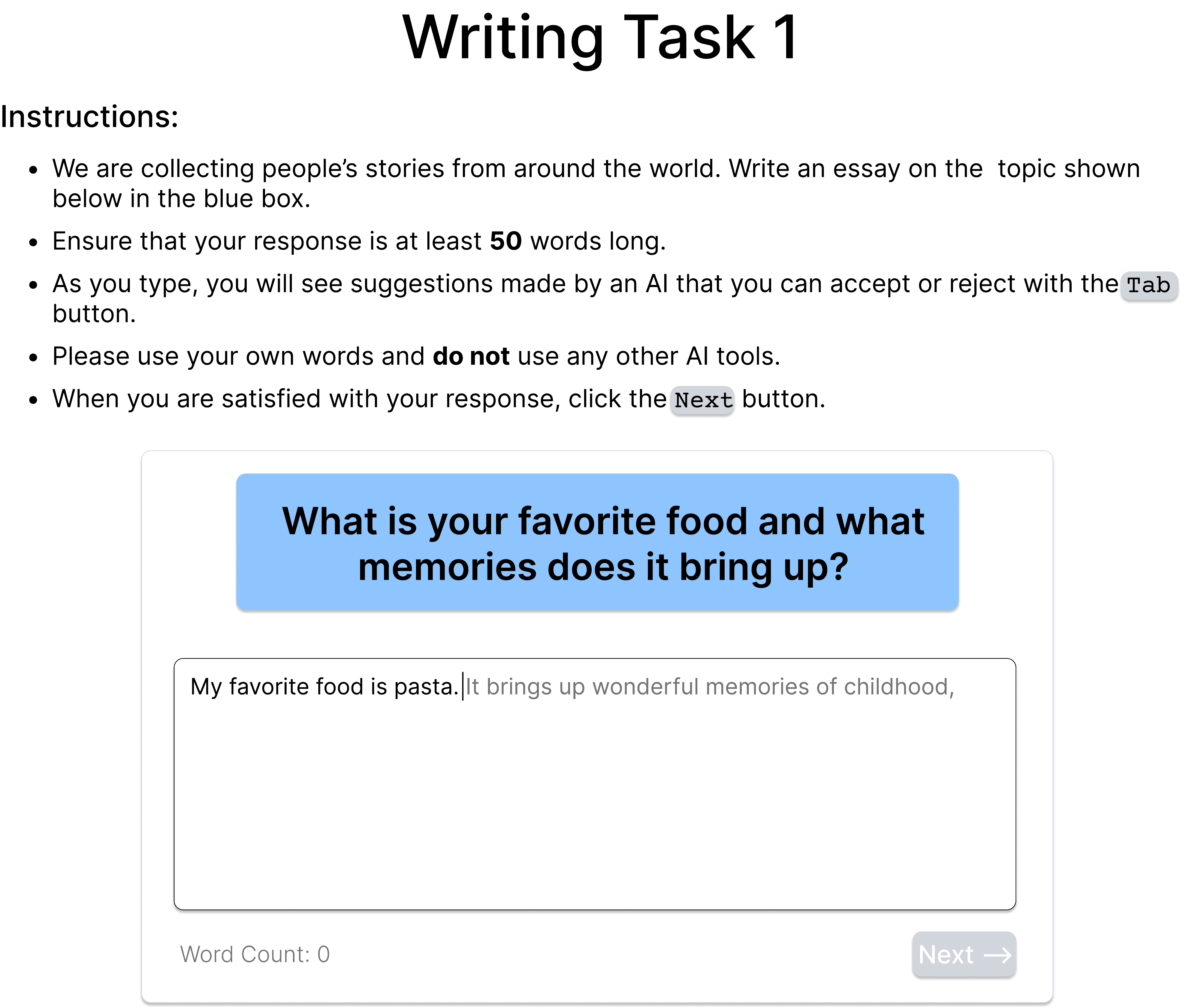}
    \caption{Our interface for the writing tasks for all participants. AI suggestions were shown in light gray and could be accepted with the \texttt{Tab} key.}
    \label{fig:writing-task-portal}
\end{figure*}

We show the interface for participants to complete our writing prompts in Fig. \ref{fig:writing-task-portal} and the full prompts in Table \ref{tab:writing-tasks}.

\begin{table*}[hbt!]
\small
\centering
    \caption{All writing prompts used for the experiment. Participants were assigned two tasks, the first writing task was randomly selected from the first column while the second task was randomly selected from the second column.}
    \begin{tabular}{p{0.45\textwidth}|p{0.45\textwidth}}
        \toprule
        \textbf{Writing Task 1} & \textbf{Writing Task 2} \\
        \midrule
         \textbf{1A.} What is your favorite food and what memories does it bring up? & \textbf{2A.} Who is your favorite celebrity or public figure and why? What was the first thing they did that made you like them? \\ 
         \textbf{1B.} What is your favorite festival/holiday and how do you celebrate it? & \textbf{2B.} Write an email to your boss asking them for a two-week leave with information about why you need to be away. \\ 
         \bottomrule
    \end{tabular}
\end{table*}\label{tab:writing-tasks}
\clearpage
\subsection{AI Literacy Survey}\label{ai-literacy}
\begin{table*}[!htbp]
    \small
    \centering
    \caption{AI Literacy questions and their corresponding dimensions. All questions were asked on a 7-point Likert scale where 1 was strongly disagree and 7 was strongly agree.} 
    \begin{tabular}{p{0.75\textwidth}p{0.15\textwidth}}
    \toprule
        \textbf{Question} & \textbf{Dimension} \\
        \midrule
        1. I can identify the AI technology employed in the applications and products I use. & Awareness \\ 
        \noalign{\smallskip}
        \hline
         \noalign{\smallskip}
        2. I frequently use AI technology to help me for work related tasks (e.g. email templates, spreadsheet analysis, report summarization). & Usage \\ 
        3. I frequently use AI technology to help me for non-work related tasks (e.g. project planning, writing poems). & Usage \\ 
        \noalign{\smallskip}
        \hline
        \noalign{\smallskip}
        4. I am aware of the limitations and shortcomings of AI technology & Ethics \\ 
        \bottomrule
    \end{tabular}
\end{table*}

%% file: files/91_demographics.tex
\section{Participant Demographics}\label{demographics}
\begin{table*}[!htpb]
    \centering
    \caption{Participant demographics from our demographic survey, grouped by country and condition}
    \scriptsize
    \begin{tabular}{lp{.20\textwidth}cccccccc}
    \toprule
          & & \multicolumn{2}{l}{\textbf{No Intervention}} & \multicolumn{2}{l}{\textbf{AI Values}} & \multicolumn{2}{l}{\textbf{Primed + AI Values}} & \multicolumn{2}{l}{\textbf{Primed + Comparative AI Values}}\\  
          &  & IND & USA & IND & USA& IND & USA& IND & USA \\
          \noalign{\smallskip}
          \cmidrule(lr){3-4} \cmidrule(lr){5-6} \cmidrule(lr){7-8} \cmidrule(lr){9-10}
          & \multicolumn{1}{r|}{n = 149} & 20 & 20 & 20 & 19 & 16 & 18 & 17 & 19 \\ 
          \noalign{\smallskip}
          \midrule
          \textbf{Gender} & \multicolumn{1}{l|}{}& \multicolumn{8}{l}{} \\ 
          & \multicolumn{1}{r|}{Female} & 6 & 13 & 9 & 12 & 4 & 12  & 9 & 11 \\ 
          & \multicolumn{1}{r|}{Male} & 14 & 7 & 11 & 7  & 11 & 6 & 7 & 8 \\
          & \multicolumn{1}{r|}{Other} & --- & --- & --- & --- & 1 & --- & 1 & --- \\
          \noalign{\smallskip}
          \midrule
          \textbf{Education} & \multicolumn{1}{l|}{}& \multicolumn{8}{l}{} \\ 
          & \multicolumn{1}{r|}{High School}  & --- & 4 & --- & 2 & --- & 1 & 2 & 4 \\ 
          & \multicolumn{1}{r|}{Some university} & --- & 8 & --- & 5 & 1 & 6 & --- & 2 \\ 
          & \multicolumn{1}{r|}{Complete university} & 11 & 4 & 9 & 11 & 9 & 8 & 7 & 11  \\ 
          & \multicolumn{1}{r|}{Some graduate or professional school}  & --- & --- & --- & --- & 1 & --- & 1 & ---  \\ 
          & \multicolumn{1}{r|}{Complete graduate or professional school} &  9 & 4 & 11 & 1 & 5 & 3 & 7 & 4 \\ 
          \noalign{\smallskip}   
          \midrule
          \textbf{Age Range} & \multicolumn{1}{l|}{}& \multicolumn{8}{l}{} \\ 
          & \multicolumn{1}{r|}{18-24} & 4 & 1 & 3 & --- & 7 & 1& 7 & 1\\ 
          & \multicolumn{1}{r|}{25-34} & 11 & 4 & 9 & 4 & 3 & 2 & 4 & 9 \\ 
          & \multicolumn{1}{r|}{35-44} & 3 & 6 & 6 & 6 & 3 & 4 & 2 & 3  \\ 
          & \multicolumn{1}{r|}{45-54} & 2 & 5 & 2 & 4 & 2 & 5  & 4 & 3 \\ 
          & \multicolumn{1}{r|}{55-64} & --- & 4 & --- & 4 & 1 & 2 & --- & 3 \\ 
          & \multicolumn{1}{r|}{65+} & --- & --- & --- & 1 & --- & 4 & --- & ---  \\ 
        \bottomrule
    \end{tabular}
    \label{tab:demographics}
\end{table*}
\clearpage

%% file: files/92_results.tex
\section{Results}\label{appendix:results}
\begin{table}[!htbp]
    \footnotesize 
    \centering
    \caption{Summary statistics for our AI reliance metrics (AI reliance, AI acceptance rate, and AI suggestion modification rate) grouped by condition.}
    \begin{tabular}{lcccccc}
        \toprule 
         & \multicolumn{2}{c}{AI Reliance} & \multicolumn{2}{c}{AI Acceptance Rate} & \multicolumn{2}{c}{AI Suggestion Modification Rate} \\
          \cmidrule(lr){2-3} \cmidrule(lr){4-5} \cmidrule(lr){6-7} 
         Condition & $M$ & $SD$ & $M$ & $SD$ & $M$ & $SD$ \\ 
         \midrule
         No Intervention & 0.364 & 0.317 & 0.087 & 0.115 & 0.114 & 0.199 \\ 
         AI Values & 0.238 & 0.309  & 0.077 & 0.148 & 0.048 & 0.174 \\ 
         Primed + AI Values & 0.297 & 0.351 & 0.065 & 0.101 & 0.042 & 0.129 \\ 
         Primed + Comparative AI Values & 0.322 & 0.375 &  0.065 & 0.089  & 0.068 & 0.140\\ 
         \bottomrule
    \end{tabular}
    \label{tab:ai-engagement-descrp}
\end{table}

\begin{table}[!htbp]
    \scriptsize
    \centering
    \caption{Results from our Bonferroni-corrected t-tests for our first set of hypotheses (\textbf{H1a, H1b}) regarding AI reliance metrics. The first three rows are the results from the t-tests between each intervention condition and the baseline condition. The last two rows are the results from the t-tests between the comparative condition and non-comparative interventions.}
    \begin{tabular}{lccccccccc}
         \toprule 
         & \multicolumn{2}{c}{AI Reliance} & \multicolumn{2}{c}{AI Acceptance Rate} & \multicolumn{2}{c}{AI Suggestion Modification Rate} \\
        \cmidrule(lr){2-3} \cmidrule(lr){4-5} \cmidrule(lr){6-7} 
         Comparison & $t(df)$ & $p$ & $t(df)$ & $p$ & $t(df)$ & $p$ \\ 
         \midrule 
         AI Values vs No intervention & \textbf{$t(156.00)=-2.537$} & \textbf{<0.001***} & $t(145.45)=-0.473$ & 0.319 & $t(154.30)=-2.235$ & 0.027 \\ 
         Primed + AI Values vs No intervention & $t(136.46)=-1.211$ & 0.114 & $t(145.88)=-1.221$ & 0.112 & $t(137.16)=-2.631$ & \textbf{<0.01**}\\ 
         Primed + Comparative AI Values vs No intervention & $t(139.78)=-0.755$ & 0.226 & $t(146.76)=-1.35$ & 0.090 & $t(142.14)=-1.665$ & 0.098\\ 
         \midrule 
         Primed + Comparative AI Values vs AI Values & $t(137.85)=1.483$ & 0.930 & $t(127.99)=-0.628$ & 0.268 & $t(142.14)=-1.665$ & 0.098\\ 
         Primed + Comparative AI Values vs Primed + AI Values & $t(137.99)=0.397$ & 0.654 & $t(133.71)=-0.044$ & 0.482 & $t(142.14)=-1.665$ & 0.098\\ 
         \bottomrule
    \end{tabular}
    \label{tab:ai-engagement}
\end{table}

\begin{table}[!htbp]
    \scriptsize
    \centering
    \caption{Summary statistics, relative change (\%), and 95\% confidence intervals for this percentage of change for our AI reliance metrics. We group the statistics by condition within our two demographic groups.}
    \begin{tabular}{lcccp{5em}cccp{5em}cccp{5em}}
    \toprule
      & \multicolumn{4}{c}{AI Reliance} & \multicolumn{4}{c}{AI Acceptance Rate} & \multicolumn{4}{c}{AI Modification Rate} \\ 
     \cmidrule(lr){2-5} \cmidrule(lr){6-9} \cmidrule(lr){10-13} 
     \textbf{IND} & $M$ & $SD$ &\% & 95\% CI  & $M$ & $SD$ &\% & 95\% CI & $M$ & $SD$ &\% & 95\% CI \\ 
     \midrule
     No intervention & 0.510 & 0.289 & & & 0.129 & 0.136 & & & & 0.130 & 0.178  \\
     AI Values & 0.357 & 0.336 & -30.1 & (-30.7, -29.6)  & 0.096 & 0.122 & -25.5 & (-26.3, -24.8) & 0.036 & 0.104 & -71.1 & (-72.1, -70.8)\\
     Primed + AI Values & 0.412 & 0.363 & -19.3 & (-20.0, -18.6) & 0.076 & 0.094 & -40.6 & (-41.5, -40.0) & 0.059 & 0.121 & -54.8 & (-56.0, -54.2)  \\
     Primed + Comparative AI Values & 0.482 & 0.413 & -5.6 & (-6.3, -4.8) & 0.098 & 0.106 & -24.1 & (-25.0, -23.4) & 0.085 & 0.153 & -34.5 & (-35.8, -33.6)  \\
     \midrule 
     \textbf{USA} & \\ 
     No intervention & 0.219 & 0.277 & & & 0.045 & 0.069 & & & 0.098 & 0.219  \\ 
     AI Values & 0.113 & 0.220 & -48.1 & (-49.2, -47.5) & 0.057 & 0.170 & 26.4 & (23.4, 29.7) & 0.058 & 0.227 & -40.8 & (-43.2, -39.3) \\ 
     Primed + AI Values & 0.195 & 0.311 & -10.6 & (-11.9, -9.4) & 0.055 & 0.106 & 22.8 & (20.9, 25.0) & 0.028 & 0.137 & -71.6 & (-73.7, -71.2) \\ 
     Primed + Comparative AI Values & 0.178 & 0.271 & -18.4 & (-19.5,-17.5) & 0.035 & 0.057 & -22.6 & (-23.8, -21.7) & 0.052 & 0.128 & -46.5 & ( -48.2, -46.0) \\ 
     \bottomrule
    \end{tabular}
    \label{tab:ai-engagement-country}
\end{table}

\begin{table}[!htbp]
    \footnotesize 
    \centering
    \caption{Summary statistics for our quantitative writing personalization metrics (lexical diversity and cosine similarity) grouped by condition.}
    \begin{tabular}{lcccc}
        \toprule 
         & \multicolumn{2}{c}{Lexical Diversity (TTR)} & \multicolumn{2}{c}{Cosine Similarity} \\
          \cmidrule(lr){2-3} \cmidrule(lr){4-5}
         Condition & $M$ & $SD$ & $M$ & $SD$  \\ 
         \midrule
         No Intervention & 0.782 & 0.057 & 0.531 & 0.161\\ 
         AI Values &  0.771 & 0.071 & 0.462 & 0.129\\ 
         Primed + AI Values & 0.784 & 0.054 & 0.581 & 0.129\\ 
         Primed + Comparative AI Values & 0.789 & 0.059 & 0.596 & 0.116\\ 
         \bottomrule
    \end{tabular}
    \label{tab:writing-metrics-descrp}
\end{table}

\begin{table}[!htbp]
    \footnotesize
    \centering
    \caption{Results from our Bonferroni-corrected t-tests for our hypotheses (\textbf{H2a, H2b}) regarding quantitative writing uniqueness metrics. The first three rows are the results from the t-tests between each intervention condition and the baseline condition. The last two rows are the results from the t-tests between the comparative values condition and non-comparative interventions.}
       \begin{tabular}{lccccc}
         \toprule 
         & \multicolumn{2}{c}{Lexical Diversity} & \multicolumn{2}{c}{Cosine Similarity} \\
        \cmidrule(lr){2-3} \cmidrule(lr){4-5}
         Comparison & $t(df)$ & $p$ & $t(df)$ & $p$ \\ 
         \midrule 
         AI Values vs No intervention & $t(147.64)=-1.075$ & 0.086 & $t(1575.13)=-10.008$ & \textbf{<0.0001}***\\ 
         Primed + AI Values vs No intervention & $t(144.60)=0.281$ & 0.390 & $t(1321.53)=6.461$ & 0.999\\ 
         Primed + Comparative AI Values vs No intervention & $t(147.25)=0.835$ & 0.020 & $t(1454.40)=8.997$ & 0.999\\ 
         \midrule 
         Primed + Comparative AI Values vs AI Values & $t(147.25)=0.835$ & 0.203 & $t(1426.45)=21.900$ & 0.999\\ 
         Primed + Comparative AI Values vs Primed + AI Values & $t(147.25)=0.835$ & 0.203 & $t(1101.74)=2.020$ & 0.978\\ 
         \bottomrule
    \end{tabular}
    \label{tab:writing-metrics}
\end{table}

%% file: files/93_addresults.tex
\section{Additional Tests}\label{sec:add-results}
Though not in our main analysis, as we were seeking to investigate whether each intervention led to decreased reliance and increased writing uniqueness and not differences between all groups, we provide the results of the ANOVA tests for fair comparison. We investigated the effects of our intervention on each of our collected metrics (AI reliance, AI acceptance rate, AI suggestion modification rate, lexical diversity, and cosine similarity).

\begin{table}[!htpb]
    \small
    \centering
    \caption{Summary of of our ANOVA results on all our collected metrics. The first three rows are metrics related to AI reliance while the last three rows are metrics related to writing personalization.}
    \begin{tabular}{l|ll}
    \toprule
         \textbf{Metric}& \textbf{$F$} & \textbf{$p$}  \\
         \midrule 
         AI Reliance & 1.916 & 0.127 \\ 
         AI Acceptance Rate &  0.634 & 0.594\\ 
         AI Suggestion Modification Rate & 3.020 & \textbf{<0.05*}\\ 
         \midrule 
         Lexical Diversity & 1.299 & 0.275 \\ 
         Cosine Similarity & 161.28 & \textbf{<0.0001***} \\ 
         \bottomrule
    \end{tabular}
    \label{tab:anova-results}
\end{table}

%% file: files/94_suggestionvals.tex
\section{Classifying Values in AI Suggestions}\label{app:suggestion-vals}
In total, participants saw 21,557 suggestions in the study and accepted 1,060 (4.92\%) of them. To investigate the overlap between the suggestions shown in the study and the framed AI values in our overview, for suggestion we classified the suggestion as (a) supporting, (b) contradicting, or (c) neutral towards the value expressed by value statement, or the question-answer pair shown from our AI value overview. For example, the suggestion \textit{``i enjoyed them with my friends''} that a participant received in the study would support the value statement \textit{``Friends is very important in my life''} from Q2 in our overview (Table \ref{tab:values survey}). We used \texttt{GPT-4o} as our classifier with \texttt{temperature=0.1, top\_p=1.0} and  the following prompt:
\begin{quote}
\begin{small}
\begin{verbatim}
Classify each autocomplete suggestion against the 15 value statements below.

For each statement, assign one label:
- "supports": the suggestion clearly aligns with or promotes the statement
- "contradicts": the suggestion clearly opposes or undermines the statement
- "neutral": no clear relationship or insufficient information

Value Statements: 
1. Family is very important in my life
2. Friends are very important in my life
3. Leisure time is rather important in my life
4. Politics is not important at all in my life
5. Work is very important in my life
6. Religion is not very important in my life
7. Children should be encouraged to learn respect for other people, feeling of responsibility, 
and tolerance.
8. If less importance is placed on work in our lives in the near future it is a good thing.
9. If more emphasis on the development of technology occurs in our life that is a good thing.
10. If there is greater respect for authority in occurs in our life I do not mind.
11. I somewhat trust people I meet for the first time.
12. Between freedom and security, freedom is more important.
13. On a scale of 1 ("not at all important") and 10 ("very important"), God's importance 
in my life is a 7.
14. I only attend religious services once a year these days apart from weddings and funerals.
15. I agree most with the basic meaning of religion is to do good to other people.

Return only a JSON array.
Use this schema for each suggestion:
{{
    "suggestion": "...",
    "labels": ["supports", "contradicts", "neutral"]
}}

Constraints:
- "labels" must be a string array of exactly 15 elements.
- Each element must be either "supports", "contradicts", or "neutral".
- The labels must correspond in order to value statements 1-15.
\end{verbatim}
\end{small}

\end{quote}

We confirmed \texttt{GPT-4o's} classification by randomly sampling 100 suggestions from the 21,557 suggestions seen in the study. The first author then identified each suggestion as supporting, contradicting, or neutral towards each of the value statements. Krippendorf's alpha was 0.668. Overall, we observed that the suggestions were largely neutral with weak support for questions in the social values dimension. Rarely did suggestions contradict the given value statement. Results are shown in Table \ref{tab:values-support}.
\begin{table}[!hptb]
    \small 
    \centering
    \caption{Percentage of suggestions that either support, contradicts, or is neutral towards the value expressed by the value statement. We observe that most values are neutral, with the most support for values being those in the social dimension.}
    \begin{tabular}{p{0.65\linewidth}|ccc}
    \toprule
    \textbf{Value} & \textbf{Supports} & \textbf{Neutral} & \textbf{Contradicts} \\ 
    \midrule
    \colorbox{blue!15}
    {1.  } Family is very important in my life & 29.94\% & 70.02\% & 0.02\%\\ 
     \colorbox{blue!15}
    {2.  } Friends are very important in my life & 11.25\% & 88.74\% & 0.01\%\\ 
    \colorbox{blue!15}
    {3.  } Leisure time is rather important in my life &  34.69\% & 65.28\% & 0.03\% \\ 
    \colorbox{blue!15}
    {4.  } Politics is not important at all in my life & 0.05\% & 99.59\% & 0.35\% \\ 
    \colorbox{blue!15}
    {5.  } Work is very important in my life & 10.37\% & 89.47\% & 0.15\% \\ 
    \colorbox{blue!15}
    {6.  } Religion is not very important in my life & 0.90\% & 98.38\% & 0.71\% \\ 
    \colorbox{blue!15}
    {7.  } Children should be encouraged to learn respect for other people, feeling of responsibility, 
    and tolerance. &  13.31\% & 86.60\% & 0.08\% \\ 
    \colorbox{blue!15}
    {8.  } If less importance is placed on work in our lives in the near future it is a good thing. & 5.67\% & 93.67\% & 0.67\% \\ 
    \colorbox{blue!15}
    {9.  } If more emphasis on the development of technology occurs in our life that is a good thing. & 1.32\% & 98.69\% & -- \\ 
    \colorbox{blue!15}
    {10.} If there is greater respect for authority in occurs in our life I do not mind. & 0.17\% & 99.81\% & 0.02\% \\ 
    \colorbox{violet!15}
    {11.} I somewhat trust people I meet for the first time. & 0.29\% & 99.67\% & 0.02\% \\ 
    \colorbox{ForestGreen!15}
    {12.} Between freedom and security, freedom is more important. & 0.43\% & 99.57\% & -- \\
    \colorbox{Dandelion!35}
    {13.} On a scale of 1 ("not at all important") and 10 ("very important"), God's importance 
    in my life is a 7. & 2.35\% & 97.65\% & -- \\
    \colorbox{Dandelion!35}
    {14.} I only attend religious services once a year these days apart from weddings and funerals. & 1.62\% & 98.11\% & 0.26\% \\ 
    \colorbox{Dandelion!35}
    {15.} I agree most with the basic meaning of religion is to do good to other people. & 14.5\% & 85.40\% & 0.09\% \\ 
\bottomrule
    \end{tabular}

    \label{tab:values-support}
\end{table}


%% file: references.bib
@inproceedings{agarwal2025, series={CHI ’25},
   title={AI Suggestions Homogenize Writing Toward Western Styles and Diminish Cultural Nuances},
   url={http://dx.doi.org/10.1145/3706598.3713564},
   DOI={10.1145/3706598.3713564},
   booktitle={Proceedings of the 2025 CHI Conference on Human Factors in Computing Systems},
   publisher={ACM},
   author={Agarwal, Dhruv and Naaman, Mor and Vashistha, Aditya},
   year={2025},
   month=apr, pages={1–21},
   collection={CHI ’25} }

@inproceedings{mun2025whyNotUseAI,
  title={Why (not) use AI? Analyzing People's Reasoning and Conditions for AI Acceptability},
  author={Mun, Jimin and Au Yeong, Wei Bin and Deng, Wesley Hanwen and Schaich Borg, Jana and Sap, Maarten},
  booktitle={AIES},
  year={2025},
  url={https://arxiv.org/abs/2502.07287}
}

@misc{wvs_wave7, 
title={World Values Survey Wave 7 (2017-2022) Cross-National Data-Set}, url={http://www.worldvaluessurvey.org/WVSDocumentationWV7.jsp}, 
author={Haerpfer, C. and Inglehart, R. and Moreno, A. and Welzel, C. and Kizilova, K. and {Diez-Medrano}, J. and Lagos, M. and Norris, P. and Ponarin, E. and Puranen, B.},
note={(eds.)},
DOI={10.14281/18241.24},  year={2024}, language={en} }

@article{lindeman2005ssvs,
author = {Marjaana Lindeman and Markku Verkasalo},
title = {Measuring Values With the Short Schwartz's Value Survey},
journal = {Journal of Personality Assessment},
volume = {85},
number = {2},
pages = {170--178},
year = {2005},
publisher = {Routledge},
doi = {10.1207/s15327752jpa8502_09},
    note ={PMID: 16171417}
}

@inproceedings{mun2024participAI,
  title={Particip-AI: A Democratic Surveying Framework for Anticipating Future AI Use Cases, Harms and Benefits},
  author={Mun, Jimin and Jiang, Liwei and Liang, Jenny and Cheong, Inyoung and DeCario, Nicole and Choi, Yejin and Kohno, Tadayoshi and Sap, Maarten},
  year={2024},
  booktitle={AIES},
  url={https://arxiv.org/abs/2403.14791}
}

@article{wang2023measure,
author = {Bingcheng Wang and Pei-Luen Patrick Rau and Tianyi Yuan},
title = {Measuring user competence in using artificial intelligence: validity and reliability of artificial intelligence literacy scale},
journal = {Behaviour \& Information Technology},
volume = {42},
number = {9},
pages = {1324--1337},
year = {2023},
publisher = {Taylor \& Francis},
doi = {10.1080/0144929X.2022.2072768},
URL = { 
        https://doi.org/10.1080/0144929X.2022.2072768
},
eprint = { 
        https://doi.org/10.1080/0144929X.2022.2072768
}
}

@inproceedings{buschek2021impact,
author = {Buschek, Daniel and Z\"{u}rn, Martin and Eiband, Malin},
title = {The Impact of Multiple Parallel Phrase Suggestions on Email Input and Composition Behaviour of Native and Non-Native English Writers},
year = {2021},
isbn = {9781450380966},
publisher = {Association for Computing Machinery},
address = {New York, NY, USA},
url = {https://doi.org/10.1145/3411764.3445372},
doi = {10.1145/3411764.3445372},
booktitle = {Proceedings of the 2021 CHI Conference on Human Factors in Computing Systems},
articleno = {732},
numpages = {13},
location = {Yokohama, Japan},
series = {CHI '21}
}

@inproceedings{kadoma2024aicomm,
author = {Kadoma, Kowe and Aubin Le Quere, Marianne and Fu, Xiyu Jenny and Munsch, Christin and Metaxa, Dana\"{e} and Naaman, Mor},
title = {The Role of Inclusion, Control, and Ownership in Workplace AI-Mediated Communication},
year = {2024},
isbn = {9798400703300},
publisher = {Association for Computing Machinery},
address = {New York, NY, USA},
url = {https://doi.org/10.1145/3613904.3642650},
doi = {10.1145/3613904.3642650},
booktitle = {Proceedings of the 2024 CHI Conference on Human Factors in Computing Systems},
articleno = {1016},
numpages = {10},
location = {Honolulu, HI, USA},
series = {CHI '24}
}

@inproceedings{dhillon2024cowriting,
author = {Dhillon, Paramveer S. and Molaei, Somayeh and Li, Jiaqi and Golub, Maximilian and Zheng, Shaochun and Robert, Lionel Peter},
title = {Shaping Human-AI Collaboration: Varied Scaffolding Levels in Co-writing with Language Models},
year = {2024},
isbn = {9798400703300},
publisher = {Association for Computing Machinery},
address = {New York, NY, USA},
url = {https://doi.org/10.1145/3613904.3642134},
doi = {10.1145/3613904.3642134},
booktitle = {Proceedings of the 2024 CHI Conference on Human Factors in Computing Systems},
articleno = {1044},
numpages = {18},
location = {Honolulu, HI, USA},
series = {CHI '24}
}

@book{baker2006glossary,
  title={Glossary of corpus linguistics},
  author={Baker, Paul},
  year={2006},
  publisher={Edinburgh University Press}
}

@article{welzel2010agency,
  title={Agency, values, and well-being: A human development model},
  author={Welzel, Christian and Inglehart, Ronald},
  journal={Social indicators research},
  volume={97},
  number={1},
  pages={43--63},
  year={2010},
  publisher={Springer}
}

@article{hohenstein2023,
author={Hohenstein, Jess
and Kizilcec, Rene F.
and DiFranzo, Dominic
and Aghajari, Zhila
and Mieczkowski, Hannah
and Levy, Karen
and Naaman, Mor
and Hancock, Jeffrey
and Jung, Malte F.},
title={Artificial intelligence in communication impacts language and social relationships},
journal={Scientific Reports},
year={2023},
month={Apr},
day={04},
volume={13},
number={1},
pages={5487},
issn={2045-2322},
doi={10.1038/s41598-023-30938-9},
url={https://doi.org/10.1038/s41598-023-30938-9}
}

@inproceedings{poddar2023influence,
author = {Poddar, Ritika and Sinha, Rashmi and Naaman, Mor and Jakesch, Maurice},
title = {AI Writing Assistants Influence Topic Choice in Self-Presentation},
year = {2023},
isbn = {9781450394222},
publisher = {Association for Computing Machinery},
address = {New York, NY, USA},
url = {https://doi.org/10.1145/3544549.3585893},
doi = {10.1145/3544549.3585893},
booktitle = {Extended Abstracts of the 2023 CHI Conference on Human Factors in Computing Systems},
articleno = {29},
numpages = {6},
location = {Hamburg, Germany},
series = {CHI EA '23}
}

@inproceedings{jakesch2023cowriting,
author = {Jakesch, Maurice and Bhat, Advait and Buschek, Daniel and Zalmanson, Lior and Naaman, Mor},
title = {Co-Writing with Opinionated Language Models Affects Users’ Views},
year = {2023},
isbn = {9781450394215},
publisher = {Association for Computing Machinery},
address = {New York, NY, USA},
url = {https://doi.org/10.1145/3544548.3581196},
doi = {10.1145/3544548.3581196},
booktitle = {Proceedings of the 2023 CHI Conference on Human Factors in Computing Systems},
articleno = {111},
numpages = {15},
location = {Hamburg, Germany},
series = {CHI '23}
}

@article{bucina2021trust,
author = {Bu\c{c}inca, Zana and Malaya, Maja Barbara and Gajos, Krzysztof Z.},
title = {To Trust or to Think: Cognitive Forcing Functions Can Reduce Overreliance on AI in AI-assisted Decision-making},
year = {2021},
issue_date = {April 2021},
publisher = {Association for Computing Machinery},
address = {New York, NY, USA},
volume = {5},
number = {CSCW1},
url = {https://doi.org/10.1145/3449287},
doi = {10.1145/3449287},
journal = {Proc. ACM Hum.-Comput. Interact.},
month = apr,
articleno = {188},
numpages = {21}
}

@article{vasconcelos2023overreliance,
author = {Vasconcelos, Helena and J\"{o}rke, Matthew and Grunde-McLaughlin, Madeleine and Gerstenberg, Tobias and Bernstein, Michael S. and Krishna, Ranjay},
title = {Explanations Can Reduce Overreliance on AI Systems During Decision-Making},
year = {2023},
issue_date = {April 2023},
publisher = {Association for Computing Machinery},
address = {New York, NY, USA},
volume = {7},
number = {CSCW1},
url = {https://doi.org/10.1145/3579605},
doi = {10.1145/3579605},
journal = {Proc. ACM Hum.-Comput. Interact.},
month = apr,
articleno = {129},
numpages = {38}
}

@article{tao2024culturalbias,
    author = {Tao, Yan and Viberg, Olga and Baker, Ryan S and Kizilcec, René F},
    title = {Cultural bias and cultural alignment of large language models},
    journal = {PNAS Nexus},
    volume = {3},
    number = {9},
    pages = {pgae346},
    year = {2024},
    month = {09},
    issn = {2752-6542},
    doi = {10.1093/pnasnexus/pgae346},
    url = {https://doi.org/10.1093/pnasnexus/pgae346},
    eprint = {https://academic.oup.com/pnasnexus/article-pdf/3/9/pgae346/59151559/pgae346.pdf},
}

@misc{santy2023nlpositionalitycharacterizingdesignbiases,
      title={NLPositionality: Characterizing Design Biases of Datasets and Models}, 
      author={Sebastin Santy and Jenny T. Liang and Ronan Le Bras and Katharina Reinecke and Maarten Sap},
      year={2023},
      eprint={2306.01943},
      archivePrefix={arXiv},
      primaryClass={cs.CL},
      url={https://arxiv.org/abs/2306.01943}, 
}

@article{palan2018prolific,
title = {Prolific.ac—A subject pool for online experiments},
journal = {Journal of Behavioral and Experimental Finance},
volume = {17},
pages = {22-27},
year = {2018},
issn = {2214-6350},
doi = {https://doi.org/10.1016/j.jbef.2017.12.004},
url = {https://www.sciencedirect.com/science/article/pii/S2214635017300989},
author = {Stefan Palan and Christian Schitter}
}

@article{braunclark06themes,
author = {Braun, Virginia and Clarke, Victoria},
year = {2006},
month = {01},
pages = {77-101},
title = {Using thematic analysis in psychology},
volume = {3},
journal = {Qualitative Research in Psychology},
doi = {10.1191/1478088706qp063oa}
}

@inproceedings{markelle2023mentalmodel,
author = {Kelly, Markelle and Kumar, Aakriti and Smyth, Padhraic and Steyvers, Mark},
title = {Capturing Humans’ Mental Models of AI: An Item Response Theory Approach},
year = {2023},
isbn = {9798400701924},
publisher = {Association for Computing Machinery},
address = {New York, NY, USA},
url = {https://doi.org/10.1145/3593013.3594111},
doi = {10.1145/3593013.3594111},
booktitle = {Proceedings of the 2023 ACM Conference on Fairness, Accountability, and Transparency},
pages = {1723–1734},
numpages = {12},
location = {Chicago, IL, USA},
series = {FAccT '23}
}

@inproceedings{bansal2019beyond,
  title={Beyond accuracy: The role of mental models in human-AI team performance},
  author={Bansal, Gagan and Nushi, Besmira and Kamar, Ece and Lasecki, Walter S and Weld, Daniel S and Horvitz, Eric},
  booktitle={Proceedings of the AAAI conference on human computation and crowdsourcing},
  volume={7},
  pages={2--11},
  year={2019}
}

@article { schwartz2006theory,
      author = "Shalom Schwartz",
      title = "A Theory of Cultural Value Orientations: Explication and Applications",
      journal = "Comparative Sociology",
      year = "2006",
      publisher = "Brill",
      address = "Leiden, The Netherlands",
      volume = "5",
      number = "2-3",
      doi = "10.1163/156913306778667357",
      pages=      "137 - 182",
      url = "https://brill.com/view/journals/coso/5/2-3/article-p137_3.xml"
}

@book{barni2009trasmettere,
  title={Trasmettere valori. Tre generazioni familiari a confronto},
  author={Barni, Daniela and others},
  year={2009},
  publisher={Unicopli}
}

@inproceedings{anderson2024homogenization,
author = {Anderson, Barrett R and Shah, Jash Hemant and Kreminski, Max},
title = {Homogenization Effects of Large Language Models on Human Creative Ideation},
year = {2024},
isbn = {9798400704857},
publisher = {Association for Computing Machinery},
address = {New York, NY, USA},
url = {https://doi.org/10.1145/3635636.3656204},
doi = {10.1145/3635636.3656204},
booktitle = {Proceedings of the 16th Conference on Creativity \& Cognition},
pages = {413–425},
numpages = {13},
location = {Chicago, IL, USA},
series = {C\&C '24}
}

@misc{johnson2022ghostmachineamericanaccent,
      title={The Ghost in the Machine has an American accent: value conflict in GPT-3}, 
      author={Rebecca L Johnson and Giada Pistilli and Natalia Menédez-González and Leslye Denisse Dias Duran and Enrico Panai and Julija Kalpokiene and Donald Jay Bertulfo},
      year={2022},
      eprint={2203.07785},
      archivePrefix={arXiv},
      primaryClass={cs.CL},
      url={https://arxiv.org/abs/2203.07785}, 
}

@book{norman1988psychology,
  title={The psychology of everyday things.},
  author={Norman, Donald A},
  year={1988},
  publisher={Basic books}
}

@inproceedings{kulesza2012mentalmodelai,
author = {Kulesza, Todd and Stumpf, Simone and Burnett, Margaret and Kwan, Irwin},
title = {Tell me more? the effects of mental model soundness on personalizing an intelligent agent},
year = {2012},
isbn = {9781450310154},
publisher = {Association for Computing Machinery},
address = {New York, NY, USA},
url = {https://doi.org/10.1145/2207676.2207678},
doi = {10.1145/2207676.2207678},
booktitle = {Proceedings of the SIGCHI Conference on Human Factors in Computing Systems},
pages = {1–10},
numpages = {10},
location = {Austin, Texas, USA},
series = {CHI '12}
}

@inproceedings{arnold2018bias,
author = {Arnold, Kenneth C. and Chauncey, Krysta and Gajos, Krzysztof Z.},
title = {Sentiment Bias in Predictive Text Recommendations Results in Biased Writing},
year = {2018},
isbn = {9780994786838},
publisher = {Canadian Human-Computer Communications Society},
address = {Waterloo, CAN},
url = {https://doi.org/10.20380/GI2018.07},
doi = {10.20380/GI2018.07},
booktitle = {Proceedings of the 44th Graphics Interface Conference},
pages = {42–49},
numpages = {8},
location = {Toronto, Canada},
series = {GI '18}
}

@inproceedings{arnold2018predictive,
author = {Arnold, Kenneth C. and Chauncey, Krysta and Gajos, Krzysztof Z.},
title = {Predictive text encourages predictable writing},
year = {2020},
isbn = {9781450371186},
publisher = {Association for Computing Machinery},
address = {New York, NY, USA},
url = {https://doi-org.offcampus.lib.washington.edu/10.1145/3377325.3377523},
doi = {10.1145/3377325.3377523},
booktitle = {Proceedings of the 25th International Conference on Intelligent User Interfaces},
pages = {128–138},
numpages = {11},
location = {Cagliari, Italy},
series = {IUI '20}
}

@inproceedings{lee2024portray,
author = {Lee, Messi H.J. and Montgomery, Jacob M. and Lai, Calvin K.},
title = {Large Language Models Portray Socially Subordinate Groups as More Homogeneous, Consistent with a Bias Observed in Humans},
year = {2024},
isbn = {9798400704505},
publisher = {Association for Computing Machinery},
address = {New York, NY, USA},
url = {https://doi.org/10.1145/3630106.3658975},
doi = {10.1145/3630106.3658975},
booktitle = {Proceedings of the 2024 ACM Conference on Fairness, Accountability, and Transparency},
pages = {1321–1340},
numpages = {20},
location = {Rio de Janeiro, Brazil},
series = {FAccT '24}
}

@inproceedings{lee2022coauthor,
author = {Lee, Mina and Liang, Percy and Yang, Qian},
title = {CoAuthor: Designing a Human-AI Collaborative Writing Dataset for Exploring Language Model Capabilities},
year = {2022},
isbn = {9781450391573},
publisher = {Association for Computing Machinery},
address = {New York, NY, USA},
url = {https://doi.org/10.1145/3491102.3502030},
doi = {10.1145/3491102.3502030},
booktitle = {Proceedings of the 2022 CHI Conference on Human Factors in Computing Systems},
articleno = {388},
numpages = {19},
location = {New Orleans, LA, USA},
series = {CHI '22}
}

@article{singh2023story,
author = {Singh, Nikhil and Bernal, Guillermo and Savchenko, Daria and Glassman, Elena L.},
title = {Where to Hide a Stolen Elephant: Leaps in Creative Writing with Multimodal Machine Intelligence},
year = {2023},
issue_date = {October 2023},
publisher = {Association for Computing Machinery},
address = {New York, NY, USA},
volume = {30},
number = {5},
issn = {1073-0516},
url = {https://doi.org/10.1145/3511599},
doi = {10.1145/3511599},
journal = {ACM Trans. Comput.-Hum. Interact.},
month = sep,
articleno = {68},
numpages = {57}
}

@inproceedings{yang2022AIAA,
  title={AI as an Active Writer: Interaction Strategies with Generated Text in Human-AI Collaborative Fiction Writing 56-65},
  author={Daijin Yang and Yanpeng Zhou and Zhiyuan Zhang and Toby Jia-Jun Li and LC Ray},
  booktitle={IUI Workshops},
  year={2022},
  url={https://api.semanticscholar.org/CorpusID:248301902}
}

@article{wagner2021wikipedia, title={It’s a Man’s Wikipedia? Assessing Gender Inequality in an Online Encyclopedia}, volume={9}, url={https://ojs.aaai.org/index.php/ICWSM/article/view/14628}, DOI={10.1609/icwsm.v9i1.14628}, abstractNote={ &lt;p&gt; Wikipedia is a community-created encyclopedia that contains information about notable people from different countries, epochs and disciplines and aims to document the world’s knowledge from a neutral point of view. However, the narrow diversity of the Wikipedia editor community has the potential to introduce systemic biases such as gender biases into the content of Wikipedia. In this paper we aim to tackle a sub problem of this larger challenge by presenting and applying a computational method for assessing gender bias on Wikipedia along multiple dimensions. We find that while women on Wikipedia are covered and featured well in many Wikipedia language editions, the way women are portrayed starkly differs from the way men are portrayed. We hope our work contributes to increasing awareness about gender biases online, and in particular to raising attention to the different levels in which gender biases can manifest themselves on the web. &lt;/p&gt; }, number={1}, journal={Proceedings of the International AAAI Conference on Web and Social Media}, author={Wagner, Claudia and Garcia, David and Jadidi, Mohsen and Strohmaier, Markus}, year={2021}, month={Aug.}, pages={454-463} }

@inproceedings{petridis2023anglekindling,
author = {Petridis, Savvas and Diakopoulos, Nicholas and Crowston, Kevin and Hansen, Mark and Henderson, Keren and Jastrzebski, Stan and Nickerson, Jeffrey V and Chilton, Lydia B},
title = {AngleKindling: Supporting Journalistic Angle Ideation with Large Language Models},
year = {2023},
isbn = {9781450394215},
publisher = {Association for Computing Machinery},
address = {New York, NY, USA},
url = {https://doi.org/10.1145/3544548.3580907},
doi = {10.1145/3544548.3580907},
booktitle = {Proceedings of the 2023 CHI Conference on Human Factors in Computing Systems},
articleno = {225},
numpages = {16},
location = {Hamburg, Germany},
series = {CHI '23}
}

@misc{gero2021sparksinspirationsciencewriting,
      title={Sparks: Inspiration for Science Writing using Language Models}, 
      author={Katy Ilonka Gero and Vivian Liu and Lydia B. Chilton},
      year={2021},
      eprint={2110.07640},
      archivePrefix={arXiv},
      primaryClass={cs.HC},
      url={https://arxiv.org/abs/2110.07640}, 
}

@inproceedings{kannan2016smartreply,
author = {Kannan, Anjuli and Kurach, Karol and Ravi, Sujith and Kaufmann, Tobias and Tomkins, Andrew and Miklos, Balint and Corrado, Greg and Lukacs, Laszlo and Ganea, Marina and Young, Peter and Ramavajjala, Vivek},
title = {Smart Reply: Automated Response Suggestion for Email},
year = {2016},
isbn = {9781450342322},
publisher = {Association for Computing Machinery},
address = {New York, NY, USA},
url = {https://doi.org/10.1145/2939672.2939801},
doi = {10.1145/2939672.2939801},
booktitle = {Proceedings of the 22nd ACM SIGKDD International Conference on Knowledge Discovery and Data Mining},
pages = {955–964},
numpages = {10},
location = {San Francisco, California, USA},
series = {KDD '16}
}

@inproceedings{bella2024tackling,
author = {Bella, G\'{a}bor and Helm, Paula and Koch, Gertraud and Giunchiglia, Fausto},
title = {Tackling Language Modelling Bias in Support of Linguistic Diversity},
year = {2024},
isbn = {9798400704505},
publisher = {Association for Computing Machinery},
address = {New York, NY, USA},
url = {https://doi.org/10.1145/3630106.3658925},
doi = {10.1145/3630106.3658925},
booktitle = {Proceedings of the 2024 ACM Conference on Fairness, Accountability, and Transparency},
pages = {562–572},
numpages = {11},
location = {Rio de Janeiro, Brazil},
series = {FAccT '24}
}

@article{chen2024conversational,
author={Chen, Kaiping
and Shao, Anqi
and Burapacheep, Jirayu
and Li, Yixuan},
title={Conversational AI and equity through assessing GPT-3's communication with diverse social groups on contentious topics},
journal={Scientific Reports},
year={2024},
month={Jan},
day={18},
volume={14},
number={1},
pages={1561},
issn={2045-2322},
doi={10.1038/s41598-024-51969-w},
url={https://doi.org/10.1038/s41598-024-51969-w}
}

@article{green2019algoinloop,
author = {Green, Ben and Chen, Yiling},
title = {The Principles and Limits of Algorithm-in-the-Loop Decision Making},
year = {2019},
issue_date = {November 2019},
publisher = {Association for Computing Machinery},
address = {New York, NY, USA},
volume = {3},
number = {CSCW},
url = {https://doi.org/10.1145/3359152},
doi = {10.1145/3359152},
journal = {Proc. ACM Hum.-Comput. Interact.},
month = nov,
articleno = {50},
numpages = {24}
}

@article{park2019slow,
author = {Park, Joon Sung and Barber, Rick and Kirlik, Alex and Karahalios, Karrie},
title = {A Slow Algorithm Improves Users' Assessments of the Algorithm's Accuracy},
year = {2019},
issue_date = {November 2019},
publisher = {Association for Computing Machinery},
address = {New York, NY, USA},
volume = {3},
number = {CSCW},
url = {https://doi.org/10.1145/3359204},
doi = {10.1145/3359204},
journal = {Proc. ACM Hum.-Comput. Interact.},
month = nov,
articleno = {102},
numpages = {15}
}

@misc{bansal2021doesexceedpartseffect,
      title={Does the Whole Exceed its Parts? The Effect of AI Explanations on Complementary Team Performance}, 
      author={Gagan Bansal and Tongshuang Wu and Joyce Zhou and Raymond Fok and Besmira Nushi and Ece Kamar and Marco Tulio Ribeiro and Daniel S. Weld},
      year={2021},
      eprint={2006.14779},
      archivePrefix={arXiv},
      primaryClass={cs.AI},
      url={https://arxiv.org/abs/2006.14779}, 
}

@article{bayati2014readmissions,
  title     = "Data-driven decisions for reducing readmissions for heart
               failure: general methodology and case study",
  author    = "Bayati, Mohsen and Braverman, Mark and Gillam, Michael and Mack,
               Karen M and Ruiz, George and Smith, Mark S and Horvitz, Eric",
  journal   = "PLoS One",
  publisher = "Public Library of Science (PLoS)",
  volume    =  9,
  number    =  10,
  pages     = "e109264",
  month     =  oct,
  year      =  2014,
  language  = "en"
}

@inproceedings{lai2020deception,
author = {Lai, Vivian and Liu, Han and Tan, Chenhao},
title = {"Why is 'Chicago' deceptive?" Towards Building Model-Driven Tutorials for Humans},
year = {2020},
isbn = {9781450367080},
publisher = {Association for Computing Machinery},
address = {New York, NY, USA},
url = {https://doi.org/10.1145/3313831.3376873},
doi = {10.1145/3313831.3376873},
booktitle = {Proceedings of the 2020 CHI Conference on Human Factors in Computing Systems},
pages = {1–13},
numpages = {13},
location = {Honolulu, HI, USA},
series = {CHI '20}
}

@misc{klingefjord2024humanvaluesalignai,
      title={What are human values, and how do we align AI to them?}, 
      author={Oliver Klingefjord and Ryan Lowe and Joe Edelman},
      year={2024},
      eprint={2404.10636},
      archivePrefix={arXiv},
      primaryClass={cs.CY},
      url={https://arxiv.org/abs/2404.10636}, 
}

@article{sorensen2024valuekaleidoscope,
   title={Value Kaleidoscope: Engaging AI with Pluralistic Human Values, Rights, and Duties},
   volume={38},
   ISSN={2159-5399},
   url={http://dx.doi.org/10.1609/aaai.v38i18.29970},
   DOI={10.1609/aaai.v38i18.29970},
   number={18},
   journal={Proceedings of the AAAI Conference on Artificial Intelligence},
   publisher={Association for the Advancement of Artificial Intelligence (AAAI)},
   author={Sorensen, Taylor and Jiang, Liwei and Hwang, Jena D. and Levine, Sydney and Pyatkin, Valentina and West, Peter and Dziri, Nouha and Lu, Ximing and Rao, Kavel and Bhagavatula, Chandra and Sap, Maarten and Tasioulas, John and Choi, Yejin},
   year={2024},
   month=mar, pages={19937–19947} }

@article{gill2008pluarlism,
 ISSN = {15336077, 17582237},
 URL = {http://www.jstor.org/stable/27749904},
 author = {Michael B. Gill and Shaun Nichols},
 journal = {Philosophical Issues},
 pages = {143--163},
 publisher = {[Wiley, Ridgeview Publishing Company]},
 title = {Sentimentalist Pluralism: Moral Psychology and Philosophical Ethics},
 urldate = {2026-01-05},
 volume = {18},
 year = {2008}
}

@article{frank2023baby,
  title={Baby steps in evaluating the capacities of large language models},
  author={Frank, Michael C},
  journal={Nature Reviews Psychology},
  volume={2},
  number={8},
  pages={451--452},
  year={2023},
  publisher={Nature Publishing Group US New York}
}

@article{
shiffrin2023probing,
author = {Richard Shiffrin  and Melanie Mitchell },
title = {Probing the psychology of AI models},
journal = {Proceedings of the National Academy of Sciences},
volume = {120},
number = {10},
pages = {e2300963120},
year = {2023},
doi = {10.1073/pnas.2300963120},
URL = {https://www.pnas.org/doi/abs/10.1073/pnas.2300963120},
eprint = {https://www.pnas.org/doi/pdf/10.1073/pnas.2300963120}}

@article{
aycinea2022norms,
author = {Diego Aycinena  and Lucas Rentschler  and Benjamin Beranek  and Jonathan F. Schulz },
title = {Social norms and dishonesty across societies},
journal = {Proceedings of the National Academy of Sciences},
volume = {119},
number = {31},
pages = {e2120138119},
year = {2022},
doi = {10.1073/pnas.2120138119},
URL = {https://www.pnas.org/doi/abs/10.1073/pnas.2120138119},
eprint = {https://www.pnas.org/doi/pdf/10.1073/pnas.2120138119}}

@inproceedings{basoah2025anthromorphic,
author = {Basoah, Jeffrey and Chechelnitsky, Daniel and Long, Tao and Reinecke, Katharina and Zerva, Chrysoula and Zhou, Kaitlyn and D\'{\i}az, Mark and Sap, Maarten},
title = {Not Like Us, Hunty: Measuring Perceptions and Behavioral Effects of Minoritized Anthropomorphic Cues in LLMs},
year = {2025},
isbn = {9798400714825},
publisher = {Association for Computing Machinery},
address = {New York, NY, USA},
url = {https://doi.org/10.1145/3715275.3732045},
doi = {10.1145/3715275.3732045},
booktitle = {Proceedings of the 2025 ACM Conference on Fairness, Accountability, and Transparency},
pages = {710–745},
numpages = {36},
location = {
},
series = {FAccT '25}
}

@article{basoah2025cscw,
author = {Basoah, Jeffrey and Cunningham, Jay L. and Adams, Erica and Bose, Alisha and Jain, Aditi and Yadav, Kaustubh and Yang, Zhengyang and Reinecke, Katharina and Rosner, Daniela},
title = {Should AI Mimic People? Understanding AI-Supported Writing Technology Among Black Users},
year = {2025},
issue_date = {November 2025},
publisher = {Association for Computing Machinery},
address = {New York, NY, USA},
volume = {9},
number = {7},
url = {https://doi.org/10.1145/3757423},
doi = {10.1145/3757423},
journal = {Proc. ACM Hum.-Comput. Interact.},
month = oct,
articleno = {CSCW242},
numpages = {51}
}

@article{northcraft1990organizational,
  title={Organizational behavior: A management challenge},
  author={Northcraft, Gregory B and Neale, Margaret Ann},
  journal={(No Title)},
  year={1990}
}

@article{simon1971human,
  title={Human problem solving: The state of the theory in 1970.},
  author={Simon, Herbert A and Newell, Allen},
  journal={American psychologist},
  volume={26},
  number={2},
  pages={145},
  year={1971},
  publisher={American Psychological Association}
}

@inproceedings{mejtoft2019designfriction,
author = {Mejtoft, Thomas and Hale, Sarah and S\"{o}derstr\"{o}m, Ulrik},
title = {Design Friction},
year = {2019},
isbn = {9781450371667},
publisher = {Association for Computing Machinery},
address = {New York, NY, USA},
url = {https://doi.org/10.1145/3335082.3335106},
doi = {10.1145/3335082.3335106},
booktitle = {Proceedings of the 31st European Conference on Cognitive Ergonomics},
pages = {41–44},
numpages = {4},
location = {BELFAST, United Kingdom},
series = {ECCE '19}
}

@inproceedings{bender2021stochastic,
author = {Bender, Emily M. and Gebru, Timnit and McMillan-Major, Angelina and Shmitchell, Shmargaret},
title = {On the Dangers of Stochastic Parrots: Can Language Models Be Too Big? },
year = {2021},
isbn = {9781450383097},
publisher = {Association for Computing Machinery},
address = {New York, NY, USA},
url = {https://doi.org/10.1145/3442188.3445922},
doi = {10.1145/3442188.3445922},
booktitle = {Proceedings of the 2021 ACM Conference on Fairness, Accountability, and Transparency},
pages = {610–623},
numpages = {14},
location = {Virtual Event, Canada},
series = {FAccT '21}
}

@book{battiste2000protecting,
  title={Protecting Indigenous knowledge and heritage: A global challenge},
  author={Battiste, Marie and Henderson, James (Sa’ke’j) Youngblood},
  year={2000},
  publisher={University of British Columbia Press}
}

@inproceedings{sap-etal-2020-recollection,
    title = "Recollection versus Imagination: Exploring Human Memory and Cognition via Neural Language Models",
    author = "Sap, Maarten  and
      Horvitz, Eric  and
      Choi, Yejin  and
      Smith, Noah A.  and
      Pennebaker, James",
    editor = "Jurafsky, Dan  and
      Chai, Joyce  and
      Schluter, Natalie  and
      Tetreault, Joel",
    booktitle = "Proceedings of the 58th Annual Meeting of the Association for Computational Linguistics",
    month = jul,
    year = "2020",
    address = "Online",
    publisher = "Association for Computational Linguistics",
    url = "https://aclanthology.org/2020.acl-main.178/",
    doi = "10.18653/v1/2020.acl-main.178",
    pages = "1970--1978"
}

@inproceedings{zhou2025relai,
  title={Rel-A.I.: An Interaction-Centered Approach To Measuring Human-LM Reliance},
  author={Zhou, Kaitlyn and Hwang, Jena D. and Ren, Xiang and Dziri, Nouha and Jurafsky, Dan and Sap, Maarten},
  year={2025},
  booktitle={NAACL},
  url={https://aclanthology.org/2025.naacl-long.556/}
}

@article{festinger1954theory,
  title={A theory of social comparison processes},
  author={Festinger, Leon},
  journal={Human relations},
  volume={7},
  number={2},
  pages={117--140},
  year={1954},
  publisher={Sage Publications Sage CA: Thousand Oaks, CA}
}

@article{fleischmann2021more,
  title={More threatening and more diagnostic: How moral comparisons differ from social comparisons.},
  author={Fleischmann, Alexandra and Lammers, Joris and Diel, Kathi and Hofmann, Wilhelm and Galinsky, Adam D},
  journal={Journal of Personality and Social Psychology},
  volume={121},
  number={5},
  pages={1057},
  year={2021},
  publisher={American Psychological Association}
}

@article{wu2012brain,
  title={Brain potentials in outcome evaluation: when social comparison takes effect},
  author={Wu, Yan and Zhang, Dexuan and Elieson, Bill and Zhou, Xiaolin},
  journal={International Journal of Psychophysiology},
  volume={85},
  number={2},
  pages={145--152},
  year={2012},
  publisher={Elsevier}
}

@article{boecker2022individuals,
  title={How individuals react emotionally to others’(mis) fortunes: A social comparison framework.},
  author={Boecker, Lea and Loschelder, David D and Topolinski, Sascha},
  journal={Journal of Personality and Social Psychology},
  volume={123},
  number={1},
  pages={55},
  year={2022},
  publisher={American Psychological Association}
}

@article{todd1994decisionaid,
title = {The Influence of Decision Aids on Choice Strategies: An Experimental Analysis of the Role of Cognitive Effort},
journal = {Organizational Behavior and Human Decision Processes},
volume = {60},
number = {1},
pages = {36-74},
year = {1994},
issn = {0749-5978},
doi = {https://doi.org/10.1006/obhd.1994.1074},
url = {https://www.sciencedirect.com/science/article/pii/S0749597884710740},
author = {Peter Todd and Izak Benbasat}
}

@inproceedings{zhang2020effect,
author = {Zhang, Yunfeng and Liao, Q. Vera and Bellamy, Rachel K. E.},
title = {Effect of confidence and explanation on accuracy and trust calibration in AI-assisted decision making},
year = {2020},
isbn = {9781450369367},
publisher = {Association for Computing Machinery},
address = {New York, NY, USA},
url = {https://doi.org/10.1145/3351095.3372852},
doi = {10.1145/3351095.3372852},
booktitle = {Proceedings of the 2020 Conference on Fairness, Accountability, and Transparency},
pages = {295–305},
numpages = {11},
location = {Barcelona, Spain},
series = {FAT* '20}
}

@inproceedings{yun2019trust,
author = {Yu, Kun and Berkovsky, Shlomo and Taib, Ronnie and Zhou, Jianlong and Chen, Fang},
title = {Do I trust my machine teammate? an investigation from perception to decision},
year = {2019},
isbn = {9781450362726},
publisher = {Association for Computing Machinery},
address = {New York, NY, USA},
url = {https://doi.org/10.1145/3301275.3302277},
doi = {10.1145/3301275.3302277},
booktitle = {Proceedings of the 24th International Conference on Intelligent User Interfaces},
pages = {460–468},
numpages = {9},
location = {Marina del Ray, California},
series = {IUI '19}
}

@inproceedings{gadiraju2023offensive,
author = {Gadiraju, Vinitha and Kane, Shaun and Dev, Sunipa and Taylor, Alex and Wang, Ding and Denton, Remi and Brewer, Robin},
title = {"I wouldn't say offensive but...": Disability-Centered Perspectives on Large Language Models},
year = {2023},
isbn = {9798400701924},
publisher = {Association for Computing Machinery},
address = {New York, NY, USA},
url = {https://doi.org/10.1145/3593013.3593989},
doi = {10.1145/3593013.3593989},
booktitle = {Proceedings of the 2023 ACM Conference on Fairness, Accountability, and Transparency},
pages = {205–216},
numpages = {12},
location = {Chicago, IL, USA},
series = {FAccT '23}
}

@inproceedings{yuxuan2025actions,
author = {Li, Yuxuan and Shirado, Hirokazu and Das, Sauvik},
title = {Actions Speak Louder than Words: Agent Decisions Reveal Implicit Biases in Language Models},
year = {2025},
isbn = {9798400714825},
publisher = {Association for Computing Machinery},
address = {New York, NY, USA},
url = {https://doi.org/10.1145/3715275.3732212},
doi = {10.1145/3715275.3732212},
booktitle = {Proceedings of the 2025 ACM Conference on Fairness, Accountability, and Transparency},
pages = {3303–3325},
numpages = {23},
location = {
},
series = {FAccT '25}
}

@article{kosslyn1989charts,
author = {Kosslyn, Stephen M.},
title = {Understanding charts and graphs},
journal = {Applied Cognitive Psychology},
volume = {3},
number = {3},
pages = {185-225},
doi = {https://doi.org/10.1002/acp.2350030302},
url = {https://onlinelibrary.wiley.com/doi/abs/10.1002/acp.2350030302},
eprint = {https://onlinelibrary.wiley.com/doi/pdf/10.1002/acp.2350030302},
year = {1989}
}

@article{gillespie2025trust,
  title={Trust, attitudes and use of artificial intelligence},
  author={Gillespie, Nicole and Lockey, Steven and Ward, Tabi and Macdade, A and Hassed, G},
  year={2025},
  publisher={University of Melbourne, KPMG}
}

@inproceedings{khan2025randomness,
author = {Khan, Ariba and Casper, Stephen and Hadfield-Menell, Dylan},
title = {Randomness, Not Representation: The Unreliability of Evaluating Cultural Alignment in LLMs},
year = {2025},
isbn = {9798400714825},
publisher = {Association for Computing Machinery},
address = {New York, NY, USA},
url = {https://doi.org/10.1145/3715275.3732147},
doi = {10.1145/3715275.3732147},
booktitle = {Proceedings of the 2025 ACM Conference on Fairness, Accountability, and Transparency},
pages = {2151–2165},
numpages = {15},
location = {
},
series = {FAccT '25}
}

@article{lee2017awareness,
  title={Awareness as a first step toward overcoming implicit bias},
  author={Lee, Cynthia},
  journal={Enhancing justice: Reducing bias},
  volume={289},
  year={2017}
}

@article{gehman2020realtoxicityprompts,
  title={Realtoxicityprompts: Evaluating neural toxic degeneration in language models},
  author={Gehman, Samuel and Gururangan, Suchin and Sap, Maarten and Choi, Yejin and Smith, Noah A},
  journal={arXiv preprint arXiv:2009.11462},
  year={2020}
}

@inproceedings{bianchi2023t2i,
author = {Bianchi, Federico and Kalluri, Pratyusha and Durmus, Esin and Ladhak, Faisal and Cheng, Myra and Nozza, Debora and Hashimoto, Tatsunori and Jurafsky, Dan and Zou, James and Caliskan, Aylin},
title = {Easily Accessible Text-to-Image Generation Amplifies Demographic Stereotypes at Large Scale},
year = {2023},
isbn = {9798400701924},
publisher = {Association for Computing Machinery},
address = {New York, NY, USA},
url = {https://doi.org/10.1145/3593013.3594095},
doi = {10.1145/3593013.3594095},
booktitle = {Proceedings of the 2023 ACM Conference on Fairness, Accountability, and Transparency},
pages = {1493–1504},
numpages = {12},
location = {Chicago, IL, USA},
series = {FAccT '23}
}

@inproceedings{ghosh2023chatgpt,
author = {Ghosh, Sourojit and Caliskan, Aylin},
title = {ChatGPT Perpetuates Gender Bias in Machine Translation and Ignores Non-Gendered Pronouns: Findings across Bengali and Five other Low-Resource Languages},
year = {2023},
isbn = {9798400702310},
publisher = {Association for Computing Machinery},
address = {New York, NY, USA},
url = {https://doi.org/10.1145/3600211.3604672},
doi = {10.1145/3600211.3604672},
booktitle = {Proceedings of the 2023 AAAI/ACM Conference on AI, Ethics, and Society},
pages = {901–912},
numpages = {12},
location = {Montr\'{e}al, QC, Canada},
series = {AIES '23}
}

@inproceedings{wang2024trust,
author = {Wang, Ruotong and Cheng, Ruijia and Ford, Denae and Zimmermann, Thomas},
title = {Investigating and Designing for Trust in AI-powered Code Generation Tools},
year = {2024},
isbn = {9798400704505},
publisher = {Association for Computing Machinery},
address = {New York, NY, USA},
url = {https://doi.org/10.1145/3630106.3658984},
doi = {10.1145/3630106.3658984},
booktitle = {Proceedings of the 2024 ACM Conference on Fairness, Accountability, and Transparency},
pages = {1475–1493},
numpages = {19},
location = {Rio de Janeiro, Brazil},
series = {FAccT '24}
}

@inproceedings{fogliato2022hai,
author = {Fogliato, Riccardo and Chappidi, Shreya and Lungren, Matthew and Fisher, Paul and Wilson, Diane and Fitzke, Michael and Parkinson, Mark and Horvitz, Eric and Inkpen, Kori and Nushi, Besmira},
title = {Who Goes First? Influences of Human-AI Workflow on Decision Making in Clinical Imaging},
year = {2022},
isbn = {9781450393522},
publisher = {Association for Computing Machinery},
address = {New York, NY, USA},
url = {https://doi.org/10.1145/3531146.3533193},
doi = {10.1145/3531146.3533193},
booktitle = {Proceedings of the 2022 ACM Conference on Fairness, Accountability, and Transparency},
pages = {1362–1374},
numpages = {13},
location = {Seoul, Republic of Korea},
series = {FAccT '22}
}

@article{guynn2015google,
  title={Google’s “bias busting” workshops target hidden prejudices},
  author={Guynn, Jessica},
  journal={USA Today},
  volume={12},
  year={2015}
}

@article{pope2018awareness,
  title={Awareness reduces racial bias},
  author={Pope, Devin G and Price, Joseph and Wolfers, Justin},
  journal={Management Science},
  volume={64},
  number={11},
  pages={4988--4995},
  year={2018},
  publisher={INFORMS}
}

@article{banerjee2018interpretation,
  title={On the interpretation of World Values Survey trust question-global expectations vs. local beliefs},
  author={Banerjee, Ritwik},
  journal={European Journal of Political Economy},
  volume={55},
  pages={491--510},
  year={2018},
  publisher={Elsevier}
}

@article{bomhoff2012easiadiff,
author = {Eduard J. Bomhoff and Mary Man-Li Gu},
title ={East Asia Remains Different: A Comment on the Index of “Self-Expression Values,” by Inglehart and Welzel},
journal = {Journal of Cross-Cultural Psychology},
volume = {43},
number = {3},
pages = {373-383},
year = {2012},
doi = {10.1177/0022022111435096},
URL = { 
        https://doi.org/10.1177/0022022111435096

},
eprint = { 
    
        https://doi.org/10.1177/0022022111435096
}
}

@article{springer2020progressive,
author = {Springer, Aaron and Whittaker, Steve},
title = {Progressive Disclosure: When, Why, and How Do Users Want Algorithmic Transparency Information?},
year = {2020},
issue_date = {December 2020},
publisher = {Association for Computing Machinery},
address = {New York, NY, USA},
volume = {10},
number = {4},
issn = {2160-6455},
url = {https://doi.org/10.1145/3374218},
doi = {10.1145/3374218},
journal = {ACM Trans. Interact. Intell. Syst.},
month = oct,
articleno = {29},
numpages = {32}
}

@article{muralidhar2025selectivetransparency,
title = {Operationalizing selective transparency using progressive disclosure in artificial intelligence clinical diagnosis systems},
journal = {International Journal of Human-Computer Studies},
volume = {204},
pages = {103591},
year = {2025},
issn = {1071-5819},
doi = {https://doi.org/10.1016/j.ijhcs.2025.103591},
url = {https://www.sciencedirect.com/science/article/pii/S107158192500148X},
author = {Deepa Muralidhar and Rafik Belloum and Ashwin Ashok}
}

@article{verplanken2002motivated,
  title={Motivated decision making: effects of activation and self-centrality of values on choices and behavior.},
  author={Verplanken, Bas and Holland, Rob W},
  journal={Journal of personality and social psychology},
  volume={82},
  number={3},
  pages={434},
  year={2002},
  publisher={American Psychological Association}
}

@ARTICLE{russo2022values,
  title    = "Changing Personal Values through {Value-Manipulation} Tasks: A
              Systematic Literature Review Based on Schwartz's Theory of Basic
              Human Values",
  author   = "Russo, Claudia and Danioni, Francesca and Zagrean, Ioana and
              Barni, Daniela",
  journal  = "Eur J Investig Health Psychol Educ",
  volume   =  12,
  number   =  7,
  pages    = "692--715",
  month    =  jun,
  year     =  2022,
  address  = "Switzerland",
  language = "en"
}

@article{sweeny2012say,
  title={Do as I say (not as I do): Inconsistency between behavior and values},
  author={Sweeny, Kate and Shepperd, James A and Howell, Jennifer L},
  journal={Basic and applied social psychology},
  volume={34},
  number={2},
  pages={128--135},
  year={2012},
  publisher={Taylor \& Francis}
}

@article{vermeir2006,
author={Vermeir, Iris
and Verbeke, Wim},
title={Sustainable Food Consumption: Exploring the Consumer ``Attitude -- Behavioral Intention'' Gap},
journal={Journal of Agricultural and Environmental Ethics},
year={2006},
month={Apr},
day={01},
volume={19},
number={2},
pages={169-194},
issn={1573-322X},
doi={10.1007/s10806-005-5485-3},
url={https://doi.org/10.1007/s10806-005-5485-3}
}

@misc{chen2026presentinglargelanguagemodels,
      title={Presenting Large Language Models as Companions Affects What Mental Capacities People Attribute to Them}, 
      author={Allison Chen and Sunnie S. Y. Kim and Angel Franyutti and Amaya Dharmasiri and Kushin Mukherjee and Olga Russakovsky and Judith E. Fan},
      year={2026},
      eprint={2510.18039},
      archivePrefix={arXiv},
      primaryClass={cs.HC},
      url={https://arxiv.org/abs/2510.18039}, 
}

@misc{rathi2025humansoverrelyoverconfidentlanguage,
      title={Humans overrely on overconfident language models, across languages}, 
      author={Neil Rathi and Dan Jurafsky and Kaitlyn Zhou},
      year={2025},
      eprint={2507.06306},
      archivePrefix={arXiv},
      primaryClass={cs.CL},
      url={https://arxiv.org/abs/2507.06306}, 
}

@misc{moore2024largelanguagemodelsconsistent,
      title={Are Large Language Models Consistent over Value-laden Questions?}, 
      author={Jared Moore and Tanvi Deshpande and Diyi Yang},
      year={2024},
      eprint={2407.02996},
      archivePrefix={arXiv},
      primaryClass={cs.CL},
      url={https://arxiv.org/abs/2407.02996}, 
}

@inproceedings{khan2025random,
author = {Khan, Ariba and Casper, Stephen and Hadfield-Menell, Dylan},
title = {Randomness, Not Representation: The Unreliability of Evaluating Cultural Alignment in LLMs},
year = {2025},
isbn = {9798400714825},
publisher = {Association for Computing Machinery},
address = {New York, NY, USA},
url = {https://doi.org/10.1145/3715275.3732147},
doi = {10.1145/3715275.3732147},
booktitle = {Proceedings of the 2025 ACM Conference on Fairness, Accountability, and Transparency},
pages = {2151–2165},
numpages = {15},
location = {
},
series = {FAccT '25}
}

@misc{rottger2024politicalcompassspinningarrow,
      title={Political Compass or Spinning Arrow? Towards More Meaningful Evaluations for Values and Opinions in Large Language Models}, 
      author={Paul Röttger and Valentin Hofmann and Valentina Pyatkin and Musashi Hinck and Hannah Rose Kirk and Hinrich Schütze and Dirk Hovy},
      year={2024},
      eprint={2402.16786},
      archivePrefix={arXiv},
      primaryClass={cs.CL},
      url={https://arxiv.org/abs/2402.16786}, 
}
